\documentclass[aps,pra,reprint,showpacs,floatfix]{revtex4-1}

\usepackage{amsmath}
\usepackage{txfonts}
\usepackage{microtype}
\usepackage{graphicx}
\usepackage{braket}
\usepackage[
breaklinks=true,
pdfauthor={Enderalp Yakaboylu, Michael Klaiber, Heiko Bauke, Karen Z.
  Hatsagortsyan, Christoph H. Keitel},
pdftitle={Relativistic features and time delay of laser-induced
  tunnel-ionization}
]{hyperref}
\usepackage{color}
\usepackage{isomath}
\renewcommand{\vec}{\vectorsym}

\newcommand*{\del}{\partial}

\DeclareMathOperator{\RE}{Re}

\DeclareMathOperator{\Ai}{Ai}
\DeclareMathOperator{\Bi}{Bi}

\graphicspath{{figures/}}

\begin{document}


\title{Relativistic features and time delay of laser-induced
  tunnel-ionization} 
\author{Enderalp Yakaboylu}
\author{Michael Klaiber}
\author{Heiko Bauke}
\author{Karen Z. Hatsagortsyan}
\author{Christoph H. Keitel}
\affiliation{Max-Planck-Institut f{\"u}r Kernphysik, Saupfercheckweg~1,
  69117~Heidelberg, Germany}

\date{\today}

\begin{abstract}
  The electron dynamics in the classically forbidden region during
  relativistic tunnel-ionization is investigated. The classical
  forbidden region in the relativistic regime is identified by defining
  a gauge invariant total energy operator. Introducing position
  dependent energy levels inside the tunneling barrier, we demonstrate
  that the relativistic tunnel-ionization can be well described by a
  one-dimensional intuitive picture. This picture predicts that, in
  contrast to the well-known nonrelativistic regime, the ionized
  electron wave packet arises with a momentum shift along the laser's
  propagation direction.  This is compatible with results from a strong
  field approximation calculation where the binding potential is assumed
  to be zero-ranged.  Further, the tunneling time delay, stemming from
  Wigner's definition, is investigated for model configurations of
  tunneling and compared with results obtained from the exact
  propagator. By adapting Wigner's time delay definition to the ionization process, the tunneling
  time is investigated in the deep-tunneling and in the
  near-threshold-tunneling regimes. It is shown that while in the
  deep-tunneling regime signatures of the tunneling time delay are not
  measurable at remote distance, it is detectable, however, in the
  latter regime.
\end{abstract}


\maketitle

\section{Introduction}

The investigation of relativistic regimes of laser-atom interactions, in
particular the strong field ionization of highly charged ions
\cite{Moore_1999,Chowdhury_2001,Dammasch_2001,Yamakawa_2003,Gubbini_2005,
  DiChiara_2008,Palaniyappan_2008,DiChiara_2010}, is feasible with
current laser technology \cite{Yanovsky_2008,RMP_2012}.  Strong field
multiphoton atomic processes in the relativistic domain are governed by
three parameters \cite{Becker_2002} which can be chosen to be the
Keldysh parameter $\gamma = \omega \sqrt{2 I_p}/E_0$
\cite{Keldysh_1965}, the barrier suppression parameter $E_0 /E_a$, and
the relativistic laser field parameter $\xi=E_0/(c\omega)$, with the
ionization potential $I_p$, the atomic field $E_a=(2I_p)^{3/2}$, the
laser's electric field amplitude $E_0$, the angular frequency $\omega$,
and the speed of light $c$ (atomic units are used throughout). At small
Keldysh parameters ($\gamma\ll 1$) the laser field can be treated as
quasistatic and the ionization is in the so-called tunneling regime up
to intensities with $E_0 /E_a\lesssim 1/10$, while for higher
intensities over-the-barrier ionization dominates \cite{Augst_1989}. In
the nonrelativistic case, the quasistatic tunnel-ionization is a
well-established mechanism which is incorporated as a first step in the
well-known simple-man three-step model of strong field multiphoton
ionization \cite{Corkum_1993}. In the first step of this intuitive
picture the bound electron tunnels out through the effective potential
barrier governed by the atomic potential $V(\vec{\vec{x}})$ and the
scalar potential of the quasistatic laser field as
$V_\mathrm{barrier}(\vec x) = \vec{x} \cdot \vec{E}(t_0) + V(\vec x)$,
where $t_0$ indicates the moment of quasistatic tunneling. In
nonrelativistic settings, the effect of the magnetic field component of
the laser field can be neglected and the quasistatic laser field is
described solely in terms of the scalar potential
$\vec{x} \cdot \vec{E}(t_0)$. In the second step the ionized electron
propagates in the continuum quasiclassically and the third step is a
potential recollision of the laser driven electron with the ionic core
that will not be considered here.

When entering the relativistic regime at $\xi^2 \gtrsim 1$, however, the
laser's magnetic field modifies the second step via an induced drift
motion of the continuum electron into the laser propagation direction.
For even stronger laser fields, that is when
$I_p/c^2\sim\gamma^2\xi^2\sim 1$, the laser's magnetic field can also
not be neglected anymore during the first step. Here, the description by
a sole scalar potential $V_\mathrm{barrier}(\vec x)$ is not valid
anymore and the intuitive picture of tunneling fails.  The presence of a
vector potential which generates the associated magnetic field led to a
controversy over the effective potential barrier
\cite{Reiss:2008:Limits}. Hence we ask, can the tunneling picture be
remedied for the application in the relativistic regime and can it be
formulated in a gauge-independent form? These questions are addressed in
this paper and it is shown that for a quasistatic electromagnetic wave,
it is possible to define a total energy operator where the tunneling
barrier can be identified without ambiguity in any gauge.

A further interesting and controversial aspect of tunneling and hence
tunnel-ionization is the issue of whether the motion of the particle
under the barrier is instantaneous or not. The question of whether the
tunneling phenomenon confronts special theory of relativity has been
raised \cite{Nimtz}. The main difficulty in the definition of the
tunneling time delay is due to the lack of a well-defined time operator
in quantum mechanics. For the generic problem of the tunneling time
\cite{MacColl_1932} different definitions have been proposed and the
discussion of their relevance still continues
\cite{Eisenbud_1948,Wigner_1955,Smith_1960,Landauer_1994,Sokolovski_2008,Steiberg_2008,Ban_2010,Galapon_2012,Vona:Hinrichs:Durr:2013:What_Does_One_Measure}.
Recent interest to this problem has been renewed by a unique opportunity
offered by attosecond angular streaking techniques for measuring the
tunneling time during laser-induced tunnel-ionization
\cite{Eckle_2008a,Eckle_2008b,Pfeiffer_2012a,Landsman_2013}.  Here we
investigate the tunneling time problem for ionization in nonrelativistic
as well as in relativistic settings. For tunnel-ionization by
electromagnetic fields there is a direct relationship of the tunneling
time to the shift in coordinate space of the ionized electron wave
packet in the laser's propagation direction at the appearance in the
continuum \cite{Klaiber_2013c}.
Within the quasiclassical description, using either the
Wentzel-Kramers-Brillouin (WKB) approximation or a path integration in
the Euclidean space-time along the imaginary time axis
\cite{Perelomov_1966a,Popov_2004u,Popov_2005,Popruzhenko_2008a,Popruzhenko_2008b},
the under-the-barrier motion is instantaneous. Thus, we address the time
delay problem by going beyond the quasiclassical description.  In the
present paper, we adopt Wigner's approach to the tunneling time
\cite{Eisenbud_1948,Wigner_1955,Smith_1960} which, in simple terms,
allows to follow the peak of the tunneled wave packet. We find
conditions when a non-vanishing Wigner time delay under the barrier for
the tunnel-ionization is expected to be measurable by attosecond angular
streaking techniques.

One of the theoretical tools applied in this paper is the relativistic
strong field approximation (SFA) \cite{Reiss_1990,Reiss_1990b}.
Neglecting the atomic potential for the continuum electron and
approximating its dynamics with a Volkov state is the main approximation
of the SFA \cite{Faisal_1973,Keldysh_1965,Reiss_1980}.  Consequently,
the prediction of the SFA is much more accurate for a zero-range
potential than for a more realistic long-range potential as the Coulomb
potential.  SFA calculations for the tunnel-ionization modeled with a
zero-range potential show that there is a momentum shift along the
laser's propagation direction due to the tunneling step.  We find that
this shift can also be estimated via a WKB analysis when a Coulomb
potential is used and that it is measurable in a detector after the
laser field has been turned off.

The structure of the paper is the following.  In
Sec.~\ref{sec:parameters} the parameter domain of the relativistic
tunneling dynamics is estimated.  In Sec.~\ref{sec:gauge} gauge
invariance in quantum mechanics is discussed and the gauge independence
of the tunneling barrier is established in nonrelativistic as well as in
relativistic settings. The intuitive picture for the tunnel-ionization
is discussed in Sec.~\ref{sec:intuitive} reducing the full problem to a
one-dimensional one. In Sec.~\ref{sec:SFA} the SFA formalism is
presented and the momentum distribution at the tunnel exit is
calculated. In Sec.~\ref{sec:tunneling_time} the tunneling time delay
and its corresponding quasiclassical counterpart are investigated and in
Sec.~\ref{sec:time_tunnel_ionization} it is applied to
tunnel-ionization. Our conclusions and further remarks are given in
Sec.~\ref{conc}.

\section{Relativistic parameters}
\label{sec:parameters}

 Let us estimate the role of relativistic effects in the
  tunnel-ionization regime which is valid for $\gamma \ll 1$ and for the
  intensities up to $E_0 /E_a < 1/10$. The typical velocity of the
  electron during the under-the-barrier dynamics can be estimated from
  the bound state energy $I_p$ as $\kappa \equiv\sqrt{2 I_p}$ (for a
  hydrogenlike ion with charge $Z$ and $I_p=c^2 - \sqrt{c^4 - Z^2 c^2}$
  in the ground state it follows $\kappa \sim Z$). The nonrelativistic
  regime of tunneling is defined via $\kappa \ll c$.  This relation is
  valid for hydrogenlike ions with nuclear charge up to $Z \sim 20$
  where $\kappa / c ~ \sim 0.14$. For ions with charge $Z>20$ the
  relativistic regime is entered because the velocity during tunneling
  is not negligible anymore with respect to the speed of light. However,
  even for an extreme case of $U^{91+}$ with $Z=92$ it is $\kappa/c \sim
  0.6$, i.\,e., the dynamics is still weakly-relativistic and a
  Foldy-Wouthuysen expansion of the relativistic Hamiltonian up to order
  $(\kappa/c)^2$ is justified. The expansion yields
  \begin{multline}
    \label{eq:FW}
    H = \frac{1}{2} \left( \vec{p} + \frac{\vec{A}(\eta)}{c}
    \right)^2 -\phi(\eta) + V(\vec{x}) -
    \frac{\vec{p}^4}{8c^2} \\
     + \frac{\vec{\sigma}\cdot \vec{B}(\eta)}{2 c} +
    i \frac{\vec{\sigma}\cdot (\vec{\nabla}\times \vec{E}(\eta))}{8 c^2}
    + \frac{\vec{\sigma}\cdot (\vec{E}(\eta)\times \vec{p})}{4 c^2} +
    \frac{\vec{\nabla}\cdot \vec{E}(\eta)}{8 c^2} \, ,
  \end{multline}
  where $V(\vec{x})$ is the binding potential, the four-vector potential
  is given by $A^\mu = (\phi(\eta), \vec{A}(\eta))$ and the phase of the
  electromagnetic wave is $\eta = x^\mu k_\mu = \omega
  (t-\hat{\vec{k}}\cdot \vec{x}/c)$, with $x^\mu = (c\, t, \vec{x})$ and
  $k^\mu = \omega/c (1,\hat{\vec{k}})$ and the laser's propagation direction $\hat{\vec{k}}$.

  In the tunneling regime, the typical displacement along the laser's
  propagation direction can be estimated as $\hat{\vec{k}}\cdot \vec{x}
  \sim \hat{\vec{k}}\cdot \vec{F}_L \, \tau_K^2$ with the Lorentz force
  $\vec{F}_L$ and the typical ionization time (Keldysh time) $\tau_K =
  \gamma /\omega =\kappa / E_0$. Identifying the Lorentz force along the
  laser's propagation direction as $\hat{\vec{k}}\cdot \vec{F}_L \sim
  \kappa B_0 /c $ ($E_0 = B_0$), the typical distance reads
  $\hat{\vec{k}}\cdot \vec{x} \sim \gamma \kappa^2 / (\omega c)$. Hence,
  electric as well as magnetic non-dipole terms are negligible since
  $\omega\,~\hat{\vec{k}}\cdot
  \vec{x}~/~c~\sim~\gamma~(\kappa/c)^2~\ll~1$, i.\,e., the typical width
  of the electron's wave packet is small compared to the laser's
  wavelength.

  Furthermore, the leading spin term in Eq.~(\ref{eq:FW}) is the
  spin-magnetic field coupling Hamiltonian $H_{P} = \vec{\sigma}\cdot
  \vec{B}/ (2 c) $. Its order of magnitude can be estimated as $H_{P} /
  \kappa^2 \sim E_0 / (c \kappa^2) = \kappa/c (E_0/ E_a)$.  Therefore,
  in the tunneling regime the spin related terms and the Darwin term
  ${\vec{\nabla}\cdot \vec{E}(\eta)}/({8 c^2})$ in Eq.~(\ref{eq:FW}) can
  be neglected because $E_0/ E_a \ll 1$. In summary, the electron's
  under-the-barrier dynamics is governed in the G{\"o}ppert-Mayer gauge, see
  Sec.~III, by the Hamiltonian
  \begin{align}
    \label{final_expansion}
    H &= H_0 + H_{ED} + H_{MD} + H_{RK} + H_{I} \, , \\
   \nonumber  & =  \frac{1}{2} \left( \vec{p} + \vec{x} \cdot
      \vec{E}(\omega t) \frac{\hat{\vec{k}}}{c} \right)^2 -
    \frac{\vec{p}^4}{8c^2} + \vec{x} \cdot \vec{E}(\omega t) +
    V(\vec{x})
  \end{align}
  with the free atomic Hamiltonian $H_0 = \vec{p}^2/2 + V(\vec{x})$, the
  electric-dipole $H_{ED} = \vec{x} \cdot \vec{E}(\omega t) $, the
  magnetic-dipole $H_{MD} = \vec{x} \cdot \vec{E}(\omega t) \vec{p}
  \cdot \hat{\vec{k}}/c$, the relativistic kinetic energy correction
  $H_{RK} = - \vec{p}^4 / (8 c^2)$, and finally $H_{I}= (\vec{x} \cdot
  \vec{E}(\omega t))^2/(2c^2)$.  For the electron's under-the-barrier
  dynamics the relative strengths of the various terms of the
  Hamiltonian (\ref{final_expansion}) are $H_{ED}/H_0 \sim 1$,
  $H_{MD}/H_0 \sim (\kappa/c)^2$, $H_{RK}/H_0 \sim (\kappa/c)^2$, and
  $H_{I}/H_0 \sim (\kappa/c)^2$ for typical displacements $\vec{x} \cdot
  \vec{E} \sim \kappa^2 $ along the polarization direction.

  In the reminder of the article, we consider tunnel-ionization from a
  Coulomb potential or zero-range potential by a monochromatic plane
  wave in the infrared ($\omega=0.05$ a.u.) and we use two extreme but
  feasible sets of parameters which ensure that we are in the tunneling
  regime, viz.\ $\kappa=90$ and $E_0/E_a=1/30$ for the deep-tunneling
  regime and $E_0/E_a=1/17$ for the near-threshold-tunneling regime.
  Formal definitions of these two regimes will be given in
  Sec.~\ref{ssc:intuitive_time_delay}.


\section{Electrodynamics, gauge freedom, and tunneling}
\label{sec:gauge}

In this section we demonstrate how the tunneling barrier in the presence
of electromagnetic fields can be defined in a gauge invariant
manner. For this purpose, we briefly summarize the gauge theory in the
light of \cite{Mills_1989,Weinberg_qm}.

\subsection{Gauge theory}

In classical electrodynamics the Maxwell equations allow to express the
physical quantities, the electric and magnetic fields, in terms of a
scalar potential $\phi$ and a vector potential $\vec{A}$
\begin{align}
  \label{e_field} \vec{E} & =
  - \vec{\nabla} \phi - \dfrac{1}{c}\del_t \vec{A}\,, \\
  \label{b_field} \vec{B} & =
  \vec{\nabla} \times \vec{A}\,.
\end{align}
Gauge invariance is the feature of electrodynamics that any other pair
of a scalar and a vector potential that is related by a so-called gauge
transformation describes the same electromagnetic fields. More
precisely, the transformation
\begin{subequations}
  \label{gauge_trans}
  \begin{align}
    \label{gauge_trans_scalar}
    \phi &\rightarrow 
    \phi' = \phi - \dfrac{1}{c} \del_t \chi\,, \\
    \label{gauge_trans_vector}
    \vec{A} &\rightarrow \vec{A'} = \vec{A} + \vec{\nabla} \chi
  \end{align}
\end{subequations}
induced via the gauge function $\chi$ leaves the electric field and the
magnetic field invariant.  Consequently, all physically measurable
electrodynamic quantities, the Maxwell equations, and the Lorentz force
law are gauge invariant.  This means they do not depend on the choice
for the electromagnetic potentials.  Furthermore, the Schr{\"o}dinger
equation for a particle in electromagnetic fields is invariant under the
transformation (\ref{gauge_trans}) provided that the state vector
transforms with the gauge transformation $U = \exp\left(-i \chi /c
\right)$ as
\begin{equation}
  \ket{\psi} \rightarrow 
  U \ket{\psi}\,.
\end{equation}

\subsection{Gauge invariant energy operator}
\label{sec:barrier}

Besides the elegance of the gauge theory, all the physical quantities,
i.\,e., the experimental observables cannot depend on the choice of the
gauge function. Let us discuss if a physical tunneling potential barrier
can be defined in a gauge independent manner.

Each physical operator that corresponds to a measurable quantity must be
gauge independent.  For example, the canonical momentum operator
$\vec{p}$, transforms under the gauge transformation $U$ as
\begin{equation}
  \vec{p} \rightarrow U \vec{p} U^\dagger = 
  \vec{p} + \dfrac{1}{c} \nabla \chi \ne \vec{p}\,.
\end{equation}
The kinetic momentum $\vec{q}(\vec{A}) = \vec{p} + \vec{A}/c$, however,
obeys
\begin{equation}
  \vec{q}(\vec{A}) \rightarrow U \vec{q} U^\dagger = 
  \vec{p} + \vec{A'}/c = \vec{q}(\vec{A'})\,.
\end{equation}
Here, the canonical momentum $\vec{p}$ which generates the space
translation and satisfies the canonical commutation relation is not a
physical measurable quantity, it is the kinetic momentum
$\vec{q}(\vec{A})$ that is measured in the experiment. In general, any
operator that satisfies the transformation
\begin{equation}
  \label{physopcond}
  \mathcal{O} (\vec{p},\vec{x},\vec{A}, \phi) \rightarrow U \mathcal{O}
  (\vec{p},\vec{x},\vec{A},
  \phi) U^\dagger = \mathcal{O}  (\vec{p},\vec{x},\vec{A'}, \phi')
\end{equation}
is called as a physical operator. For instance, the Hamiltonian for a
charge particle interacting with an arbitrary electromagnetic field in
the nonrelativistic regime
\begin{equation}
 H = \dfrac{ \left( \vec{p}+ \vec{A}/c \right)^2}{2} - \phi
\end{equation}
transforms under the gauge transformation as
\begin{equation}
  H \rightarrow \dfrac{ \left( \vec{p}+ \vec{A'}/c \right)^2}{2} - \phi\,.
\end{equation}
Hence, it cannot be a physical operator (because $\phi$ is not equally
transformed to $\phi'$), while $H - i \del/\del t$ is the physical
operator which guarantees the invariance of the Schr{\"o}dinger equation
under a gauge transformation.

In contrast to the Hamiltonian $H$, the total energy of a system has to
be a gauge invariant physical quantity. Therefore, we have to
distinguish two concepts: the Hamiltonian and the total energy. The
Hamiltonian is the generator of the time translation, while the total
energy is defined as a conserved quantity of the dynamical system under
a time translation symmetry of the Lagrangian.  As a consequence, if the
Hamiltonian is explicitly time independent, then the Hamiltonian
coincides with the total energy operator.

For a time independent electromagnetic field there exists a certain
gauge where the Hamiltonian is explicitly time independent. The
identification of the Hamiltonian as a total energy operator implies,
then, that both the vector potential $\vec{A}$ and the scalar potential
$\phi$ associated to the constant electromagnetic field have to be time
independent. This leads to the fact that
\begin{equation}
  \phi = - \int^{\vec{x}} \vec{E}(\vec{x}') \cdot d \vec{x}'
\end{equation}
where we have used Eq.~(\ref{e_field}). In this gauge, the Hamiltonian
which coincides with the total energy operator $\hat{\varepsilon}$ in
the presence of any external potential $V(\vec x)$ reads
\begin{equation}
  \label{inital_definition_energy}
  H = \hat{\varepsilon} = 
  \dfrac{(\vec{p}+\vec{A}(\vec{x})/c)^2}{2} + \int^{\vec{x}} \vec{E}(\vec{x}')
\cdot d \vec{x}' + V(\vec x) \,,
\end{equation}
where the time independent vector potential $\vec{A}(\vec{x})$ generates
the associated magnetic field via Eq.~(\ref{b_field}).

Accordingly, if we identify Eq.~(\ref{inital_definition_energy}) as a
definition of the gauge independent total energy operator, it reads in
an arbitrary gauge
\begin{equation}
  \label{energy_operator}
  \hat{\varepsilon} = \dfrac{(\vec{p}+\vec{A}(\vec{x},t)/c)^2}{2} +
  \int^{\vec{x}} \vec{E}(\vec{x}') \cdot d \vec{x}' + V(\vec x)\,,
\end{equation}
where we have used the transformation~(\ref{physopcond}). The first term
on the right hand side of (\ref{energy_operator}) is the kinetic energy
for an arbitrary vector potential $\vec{A}(\vec{x},t)$ that appears in
the corresponding Hamiltonian.  The second term should not be regarded
as a scalar potential, but defines the potential energy. The energy
operator (\ref{energy_operator}) fulfills the conservation law
\begin{equation}
  \label{conservation}
  \dfrac{d \hat{\varepsilon}}{d t} = i \left[H, \hat{\varepsilon}\right] +
  \dfrac{\del \hat{\varepsilon}}{\del t} = 0\,,
\end{equation}
which can be prooven in a straightforward calculation.  We find
\begin{align}
  \dfrac{\del \hat{\varepsilon}}{\del t} &= 
  \dfrac{1}{2} \left( \left(\vec{p} +
      \vec{A}/c \right) \cdot \dfrac{\del \vec{A}}{c \del t} +   \dfrac{\del
      \vec{A}}{c \del t} \cdot \left(\vec{p} + \vec{A}/c \right)\right), \\
  \nonumber
  \left[H, \hat{\varepsilon}\right] &=
  \dfrac{1}{2} \left( \left(\vec{p} +
      \vec{A}/c \right) \cdot \left[\vec{p}, \int^{\vec{x}} \vec{E} \cdot d
\vec{x}'+
      \phi \right]  \right. \\ 
  &\qquad+ \left. \left[\vec{p}, \int^{\vec{x}} \vec{E} \cdot d \vec{x}' + \phi
    \right]
    \cdot \left(\vec{p} + \vec{A}/c \right) \right)
\end{align}
and hence
\begin{multline}
  \dfrac{d \hat{\varepsilon}}{d t} = \dfrac{1}{2} \left(  \left(\vec{p}
      +\dfrac{\vec{A}}{c}\right) \cdot \left( \vec{E} + \vec{\nabla}\phi +
\dfrac{\del
        \vec{A}}{c \del t} \right) \right. \\
  + \left. \left( \vec{E} + \vec{\nabla}\phi + \dfrac{\del \vec{A}}{c \del t}
    \right) \cdot \left(\vec{p} +\dfrac{\vec{A}}{c}\right) \right) = 0
\end{multline}
is obtained. The definition~(\ref{energy_operator}), then, suggests to
introduce the gauge independent effective potential energy as
\begin{equation}
  \label{barrier}
  V_\mathrm{eff}(\vec x) = 
  \int^{\vec{x}} \vec{E}(\vec{x}') \cdot d \vec{x}' +  V(\vec x) \,.
\end{equation}

In conclusion, the electron dynamics during ionization can be described
as tunneling through a potential barrier if the total energy of the
electron is conserved. Then, the potential energy and the tunneling
barrier can be identified unambiguously.  Thus, for ionization in a
laser field we have to identify the quasistatic limit such that the
tunneling picture becomes applicable.  The tunnel-ionization regime in a
laser field is determined by the Keldysh parameter $\gamma \ll 1$.  It
defines the so-called tunneling formation time $\tau_K=\gamma/\omega$,
see Sec.~\ref{formation_time_s}. The tunneling regime $\gamma \ll 1$
corresponds to situations when the formation time of the ionization
process is much smaller than the laser period. Consequently, the
electromagnetic field can be treated as quasi-static during the
tunneling ionization process and the electron energy is approximately
conserved.  Therefore, the gauge-independent operator for the total
energy $\hat{\varepsilon}$ in a quasistatic electromagnetic field can be
defined and from the latter the gauge-independent potential energy can
be deduced, which in the case of tunnel-ionization constitutes the
gauge-independent tunneling barrier.  Therefore, Eq.~(\ref{barrier})
defines the gauge independent tunneling barrier in the tunnel-ionization
regime.  In the long wavelength approximation it yields
\begin{equation}
  V_\mathrm{barrier} = \vec{x} \cdot \vec{E}(t_0)  + V(\vec x)\,,
\end{equation}
where $t_0$ is the moment of ionization and $V(\vec x)$ is the binding
potential for the tunnel-ionization.

As an illustration of the gauge independence of the tunneling barrier,
let us compare two fundamental gauges used in strong field physics to
describe nonrelativistic ionization. In the length gauge where $\phi = -
\vec{x} \cdot \vec{E}_0$, $\vec{A} = 0$, the nonrelativistic Hamiltonian
for a constant uniform electric field is given by
\begin{equation}
  H = \dfrac{\vec{p}^2}{2} + \vec{x} \cdot \vec{E}_0 + V(\vec x)\,.
\end{equation}
Here, the Hamiltonian coincides \textbf{with} the physical energy operator.  In 
the
velocity gauge, however, where $\vec{A} = - c \vec{E}_0 t$, $\phi = 0$,
the same dynamics is governed by the Hamiltonian
\begin{equation}
  \label{H_velocity}
  H = \dfrac{(\vec{p}-\vec{E}_0 t)^2}{2} + V(\vec x)\,.
\end{equation}
In Eq.~(\ref{H_velocity}) it seems as if there is no potential
barrier. However, the \textit{conserved} energy operator
\begin{equation}
  \hat{\varepsilon} = 
  \dfrac{(\vec{p}-\vec{E}_0 t)^2}{2} + \vec{x} \cdot\vec{E}_0 +
  V(\vec x),
\end{equation}
reveals the tunneling barrier $\vec{x} \cdot \vec{E}_0 + V(\vec x)$.
Thus, for arbitrary time independent (quasistatic) electromagnetic
fields, the gauge-independent tunneling barrier can be defined without
any ambiguity.

The physical energy operator and the tunneling barrier can be
generalized to the relativistic regime straightforwardly by using the
relativistic Dirac Hamiltonian
\begin{equation}
  H = c\vec{\alpha} \cdot (\vec{p}+\vec{A}/c) - 
  \phi + V(\vec x) + \beta c^2 \,.
\end{equation}
From Eq.~(\ref{energy_operator}) we deduce the physical energy
operator in the relativistic case as
\begin{equation}
  \label{energy_operator2}
  \hat{\varepsilon} = c\vec{\alpha} \cdot (\vec{p}+\vec{A}(\vec{x},t)/c) +
  \int^{\vec{x}} \vec{E}(\vec{x}') \cdot d \vec{x}' + V(\vec x)+\beta c^2\,.
\end{equation}

One possible generalization of the length gauge into the relativistic
regime is the G{\"o}ppert-Mayer gauge
\begin{equation}
  A^\mu = - \vec{x} \cdot \vec{E}(\eta) (1, \hat{\vec{k}}) \, .
  \label{GM} 
\end{equation}
Taking into account that the dipole
approximation for the laser field can be applied inside the tunneling
barrier, the Hamiltonian which coincides with the total relativistic
energy operator in the G{\"o}ppert-Mayer gauge reads
\begin{equation}
  \label{eq:energy_G_M}
  H = \hat{\varepsilon} = 
  c\vec{\alpha} \cdot
  \left(\vec{p}-\hat{\vec{k}}\frac{\vec{x}\cdot\vec{E}(\eta_0)}{c}\right)
  + \vec{x} \cdot \vec{E}(\eta_0) + V(\vec x),
\end{equation}
where $\eta_0$ is the laser phase at the moment of ionization.

\begin{figure}
  \centering
  \includegraphics[scale=0.525]{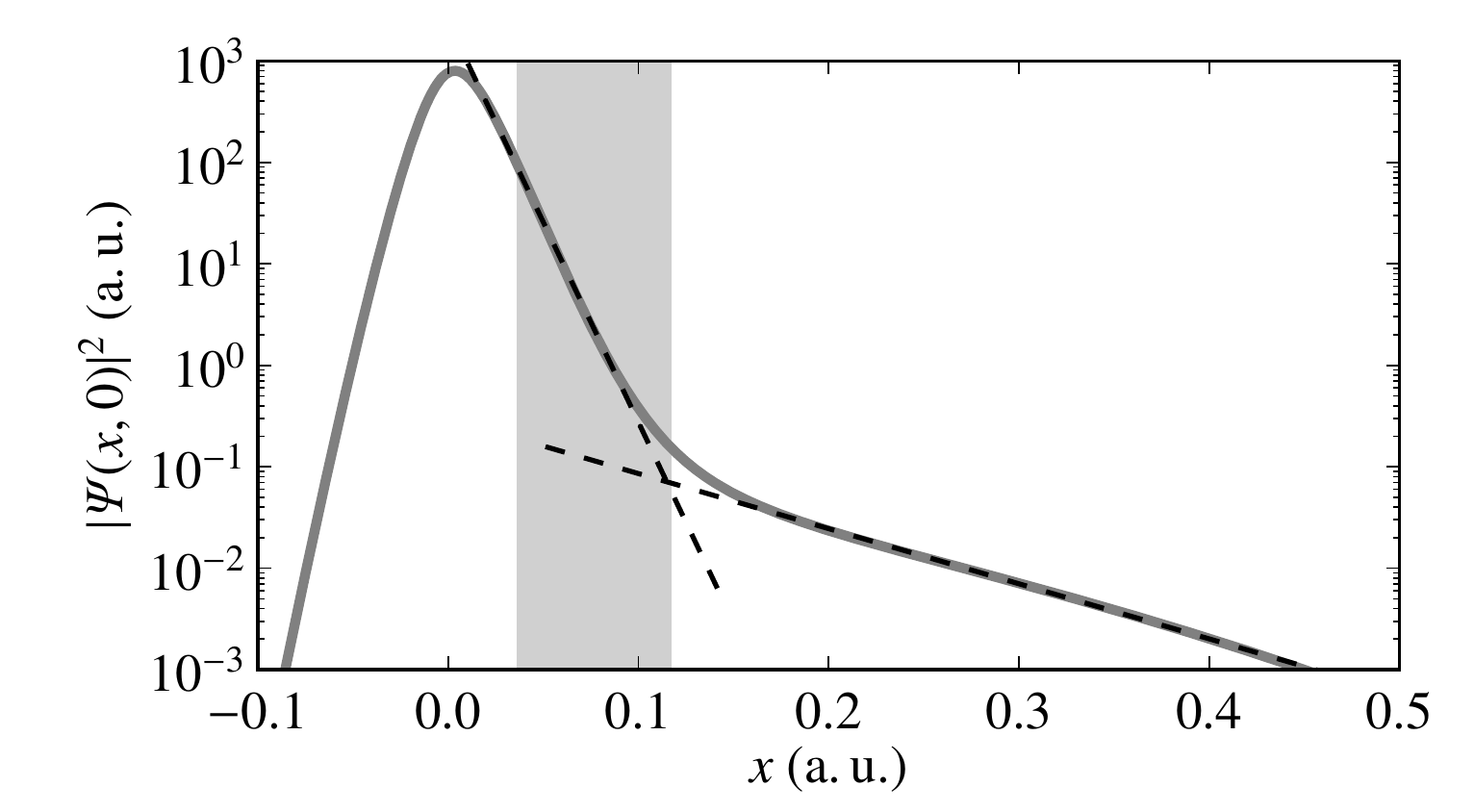}
  \caption{The electron density (solid line) along the laser
    polarization direction at the instant of maximal field strength at
    the atomic core. The shaded area represents the classical
    forbidden area, gray dashed lines are exponential fits on the wave
    function density. The electron's wave function $\Psi(x, z)$ is
    obtained by solving the two-dimensional Dirac equation for an
    electron in a soft-core potential interacting with an external
    laser pulse, laser parameters are the same as in
    \cite{Klaiber_2013c}.}
  \label{fig:wf}
\end{figure} 

The tunneling barrier results from an interpretation of the individual
mathematical terms of the quasistatic energy
operator~(\ref{eq:energy_G_M}). It has, however, also a physical
significance as it can be demonstrated by an ab initio numerical
simulation of the tunneling process in a highly charged ion in a laser
field of relativistic intensities based on the Dirac equation
\cite{Bauke_2011b,Klaiber_2013c}.  Figure~\ref{fig:wf} shows the
gauge-independent electron density along the laser polarization
direction at the instant of maximal field strength at the atomic core.
The electron density can be divided in two parts that are characterized
by two different decay rates. The switchover region includes the
tunneling exit that is defined by the tunneling barrier.  The decay of
the density under the barrier is related to damping due to tunneling,
i.\,e., approximately $\sim\exp(-\kappa x)$, whereas outside the barrier
it is dominated by transversal spreading.  As the change of slopes
occurs close to the tunneling exit the tunneling barrier is real and
physical and not just a result of an interpretation in a particular
gauge.

\section{Intuitive picture for the relativistic tunnel-ionization
  process}
\label{sec:intuitive}

Having identified the gauge invariant tunneling barrier, we elaborate in
this section on the intuitive picture for the tunnel-ionization process
in the relativistic regime.  For the reminder of the article we choose
our coordinate system such that the laser's electric field component
$\vec E_0$ is along the $x$ direction, the laser's magnetic component
$\vec B_0$ is along the $y$ direction and the laser propagates along the
$z$ direction.  Since we assume to work in the quasistatic regime, the
the G{\"o}ppert-Mayer gauge is applied for the Hamiltonian such that it
coincides with the energy operator.  In the nonrelativistic limit the
latter agrees with the Schr{\"o}dinger Hamiltonian in the length gauge.

\subsection{Nonrelativistic case}

\begin{figure}
  \centering
  \includegraphics[scale=0.6]{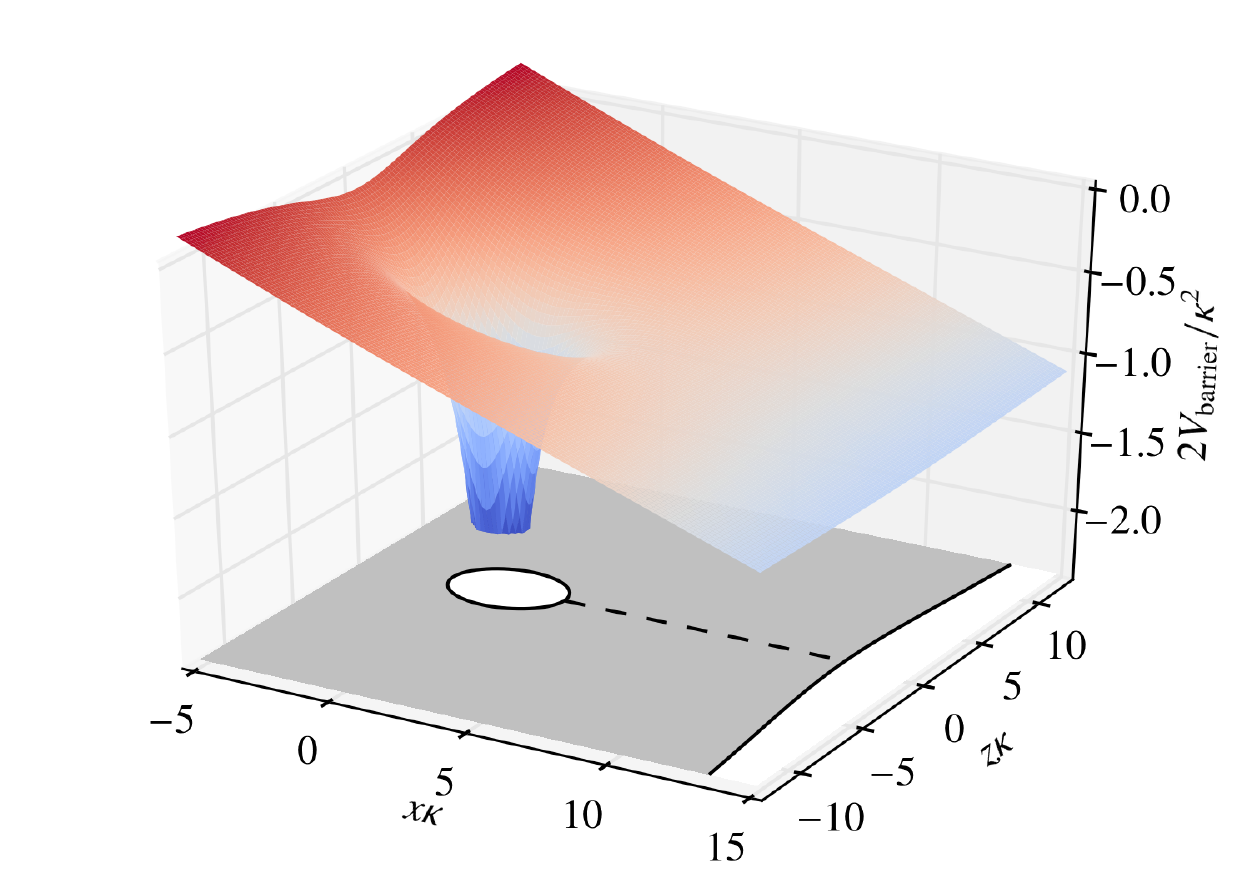}
  \caption{(Color online) The potential barrier $V_\mathrm{barrier}$ for
    quasistatic tunnel-ionization.
    The laser field $E(t_0) = - \kappa^3/30$ is along the $x$ direction.
    The most probable tunneling path is indicated by the black dashed
    line.}
  \label{fig:barrier}
\end{figure}

In the nonrelativistic limit the intuitive picture for the
tunnel-ionization is well-known.  In this picture the magnetic field and
nondipole effects can be neglected, and the Hamiltonian reads
\begin{equation}
  H = \dfrac{\vec{p}^2}{2} + x E(t_0) - \dfrac{\kappa}{r}\, ,
\end{equation}
with $r=\sqrt{x^2+y^2+z^2}$. Introducing the potential 
\begin{equation}
  V_\mathrm{barrier}(\vec x) = x E(t_0) - \dfrac{\kappa}{r},
\end{equation}
one can define the classical forbidden region.  The tunneling
probability increases with decreasing width of the barrier, thus, the
most probable tunneling path is concentrated along the electric field
direction as indicated by the dashed line in Fig.~\ref{fig:barrier}.
Therefore, it is justified to restrict the analysis of the tunneling
dynamics along the laser's polarization
direction 
\footnote{\label{1D} The one-dimensional motion along the laser electric
  field during tunneling ionization can also be justified from the
  following estimation. The role of different forces can be evaluated by
  their contribution to the action, which can be estimated by an order
  of magnitude as $S\sim \varepsilon \tau$, where $\varepsilon$ is the
  typical energy and $\tau$ is the typical time of an acting force. The
  contribution from the laser electric field is $S_L\sim \tau_K x_e
  E_0\sim\kappa^3/E_0=E_a/E_0$, with the typical distance on which the
  laser electric field acts on the tunneling electron $x_e\sim
  \kappa^2/E_0$ (the barrier length) and the Keldysh time $\tau_K$.  The
  contribution from the Coulomb potential can be separated into two
  parts. The longitudinal Coulomb force contribution is of the order of
  $S_c^\parallel \sim (\kappa/x_c) \tau_c\sim 1$, with the typical time
  $\tau_c$ and coordinate $x_c\sim \kappa \tau_c$, where the Coulomb
  force makes the main contribution on the electron. The transverse
  Coulomb force contribution can be estimated via $S_c^\bot \sim
  F_c^\bot z_c \tau_c \sim z_c^2/x_c^2\sim \sqrt{E_0/E_a}$, where the
  transverse Coulomb force is $F_c^\bot\sim \kappa z_c/x_c^3$, the
  typical longitudinal coordinate is derived equating the laser and
  Coulomb forces $\kappa/x_c^2=E_0$, and the typical transverse coordinate
  is estimated from $\kappa z_c^2\sim x_c$. Therefore, the longitudinal
  contribution of the Coulomb potential into the dynamics represents the
  leading order correction to the zero-range potential case, while the
  transversal effect of the Coulomb potential is an higher order
  correction in the tunneling regime where $E_0/\kappa^3$ is small and
  will be neglected in the following.}.

\begin{figure}
  \centering
  \includegraphics[scale=0.7]{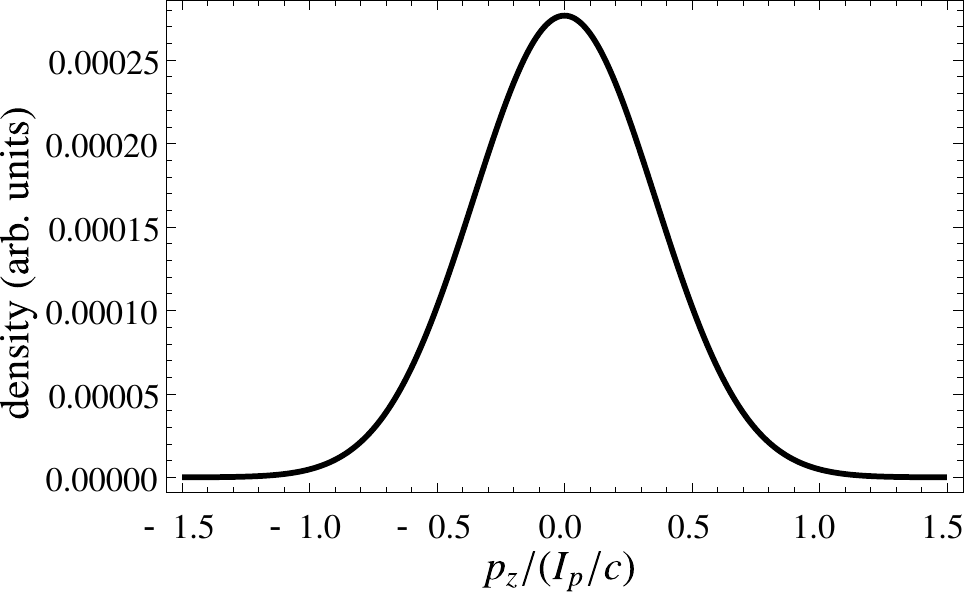}
  \caption{The nonrelativistic tunneling probability versus the momentum
    along $z$ direction. The maximum tunneling probability occurs at
    $p_z = 0$. The applied parameters are $E(t_0) = - \kappa^3/30$ and
    $\kappa = 90$. Without loss of generality, $p_y=0$ was chosen.}
  \label{non_rel_mom}
\end{figure}

\begin{figure*}
  \centering
  \includegraphics[scale=0.5]{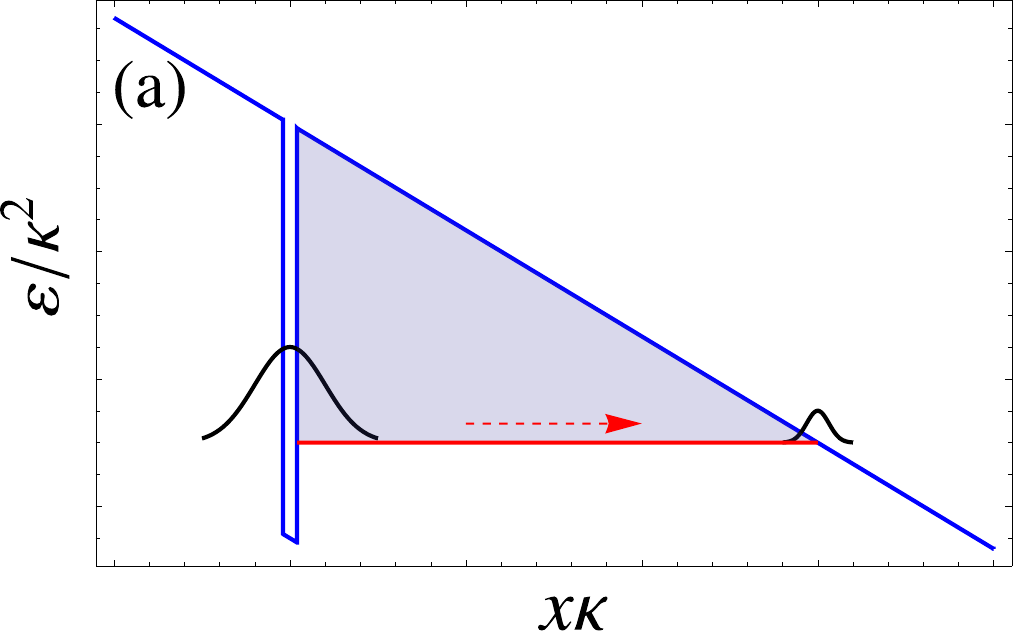}\qquad
  \includegraphics[scale=0.5]{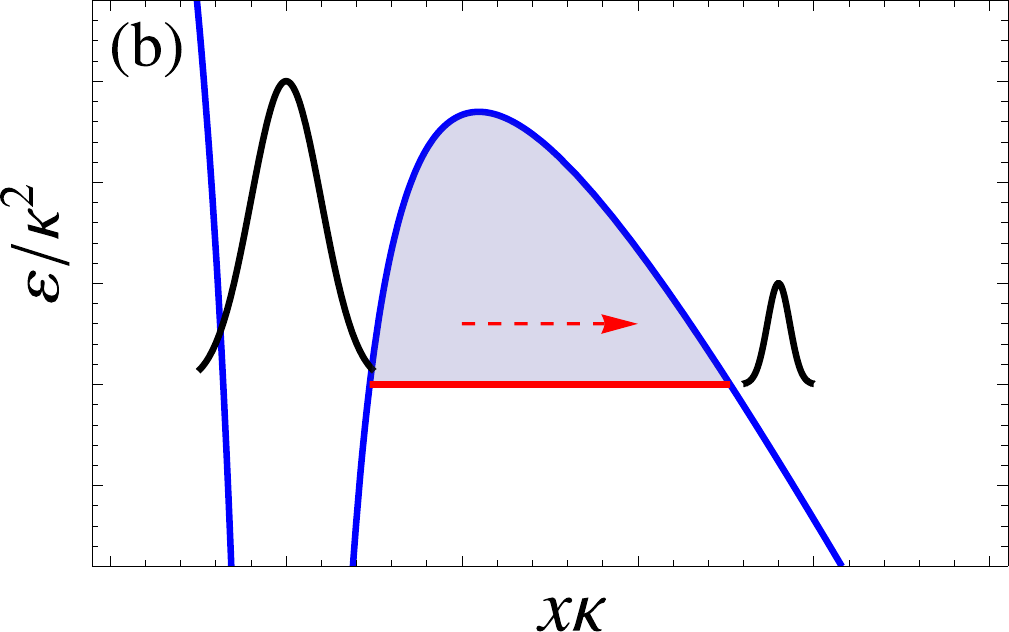}\qquad
  \includegraphics[scale=0.5]{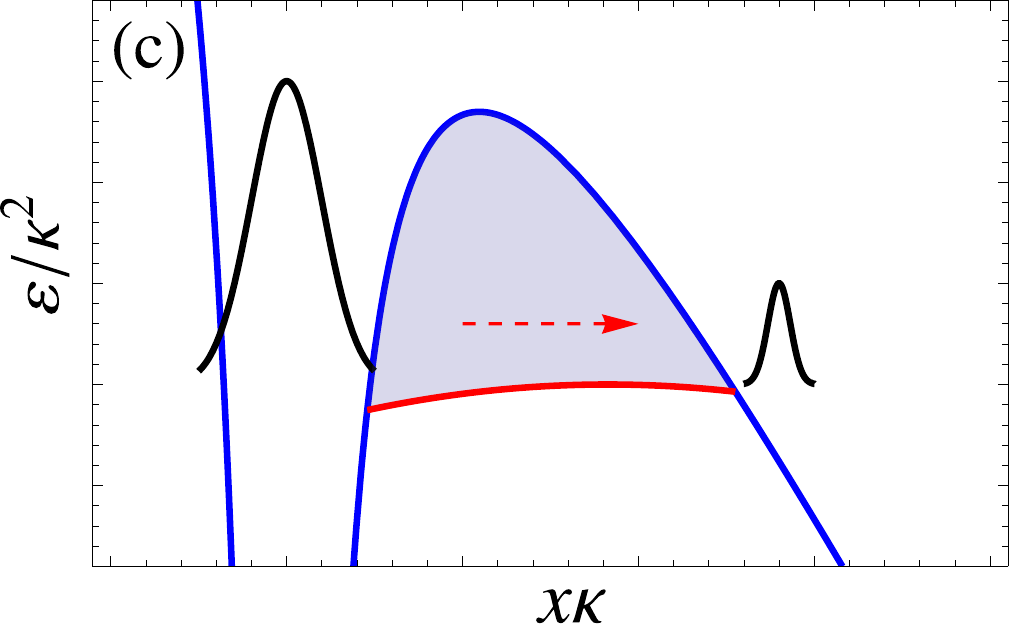}
  \caption{(Color online) Schematic picture of nonrelativistic
    (sub-figures (a) and (b)) and relativistic (subfigure (c)) tunneling
    from a bound state into the continuum: (a) the atomic potential is
    aproximated by a zero-range potential; (b) and (c) the Coulomb
    potential case.  The potential barrier (solid, blue) and the energy
    levels (dashed, red) for $p_z=0$ (in the nonrelativistic cases) and
    for the most probable transversal momentum $p_z$ (relativistic case)
    are plotted against the longitudinal tunneling coordinate $x$.  The
    shaded area can be interpreted as a measure for the tunneling
    probability. The electron wave packet is indicated in black.}
  \label{nrtp}
\end{figure*}

In this one-dimensional picture the barrier for the tunnel-ionization
follows as
\begin{equation}
  \label{tunneling_barrier_non_rel}
  V_{\mathrm{barrier}} = x  E(t_0) -\dfrac{\kappa}{|x|} \,.
\end{equation}
The momentum components $p_y$ and $p_z$ along the $y$ and the $z$
direction are conserved 
and tunneling along the $x$ direction is governed by the energy
\begin{equation}
  \label{epsilon_nr}
  \varepsilon_x = -I_p - \dfrac{p_y^2}{2}- \dfrac{p_z^2}{2}\,.
\end{equation}
The wave function of the electron and the corresponding transition
probability can be derived within the Wentzel-Kramers-Brillouin (WKB)
approximation.  The zeroth order WKB wave function is given by
\begin{equation}
  \psi \propto \exp\left(i S_{cl} \right)\,,
\end{equation}
with the classical action
\begin{equation}
  S_\mathrm{cl} = -\varepsilon_x t + \int^x p_x (x') \,dx'\,, 
\end{equation}
and the momentum's $x$ component
\begin{equation}
  \label{p_x}
  p_x (x) =
  \sqrt{2(\varepsilon_x - V_{\mathrm{barrier}})}\,.
\end{equation}
The WKB tunneling probability follows as
\begin{equation}
  |T|^2 \propto \exp\left(- 2 \int_{x_0}^{x_e} d x \, |p_x (x)| \right)\,,
  \label{exponent}
\end{equation}
where $x_0$ and $x_e$ are the entry point and exit point of the barrier
such that $p(x_0) = p(x_e)= 0$.  The dependence of the tunneling
probability on the momentum $p_z$ is shown Fig.~\ref{non_rel_mom}.  The
tunneling probability is maximal for $p_z = 0$, because the energy
level~(\ref{epsilon_nr}) decreases with increasing $p_z^2$. From this it
follows that the exit coordinate increases with increasing $p_z$.

In summary, nonrelativistic tunneling from an atomic potential can be
visualized by an one-dimensional picture given in Fig.~\ref{nrtp}(a)
and~(b). The area between the barrier and the energy level represents a
messure for the probability of the process. The larger the area the less
likely the ionization.

\subsection{Relativistic case}

In the relativistic regime, the largest correction to the
nonrelativistic Hamiltonian comes from the magnetic dipole term. Let us
consider the role of the magnetic dipole interaction in the laser field
for the tunneling picture. The corresponding time-independent
Schr{\"o}dinger equation reads
\begin{equation}
  \left[\dfrac{(-i\vec{\nabla}-x E(t_0) \hat{\vec{z}}/c)^2}{2} + x E(t_0) -
    \dfrac{\kappa}{r} \right]
  \psi(\vec{x}) = \varepsilon \, \psi(\vec{x})\,.
  \label{Schrod1}
\end{equation}
Similar to the nonrelativistic case, an approximate one-dimensional
description is valid for the most probable tunneling path along the
electric field direction. Restricting the dynamics along the electric
field direction and neglecting the dependence of the ionic core's potential
on the transverse coordinate, we have $p_{y,z}=\mathrm{const}$, and the
momentum along the polarization direction is given by Eq.~(\ref{p_x})
with the barrier~(\ref{tunneling_barrier_non_rel}), which is the same as
in the nonrelativistic case. The energy, however, is modified by the
magnetic dipole term
\begin{equation}
  \label{epsilon_r}
  \varepsilon_x =
  -I_p - \dfrac{p_y^2}{2}- \dfrac{(p_z-x E(t_0)/c)^2}{2}\,. 
\end{equation}

\begin{figure}
  \centering
  \includegraphics[scale=0.7]{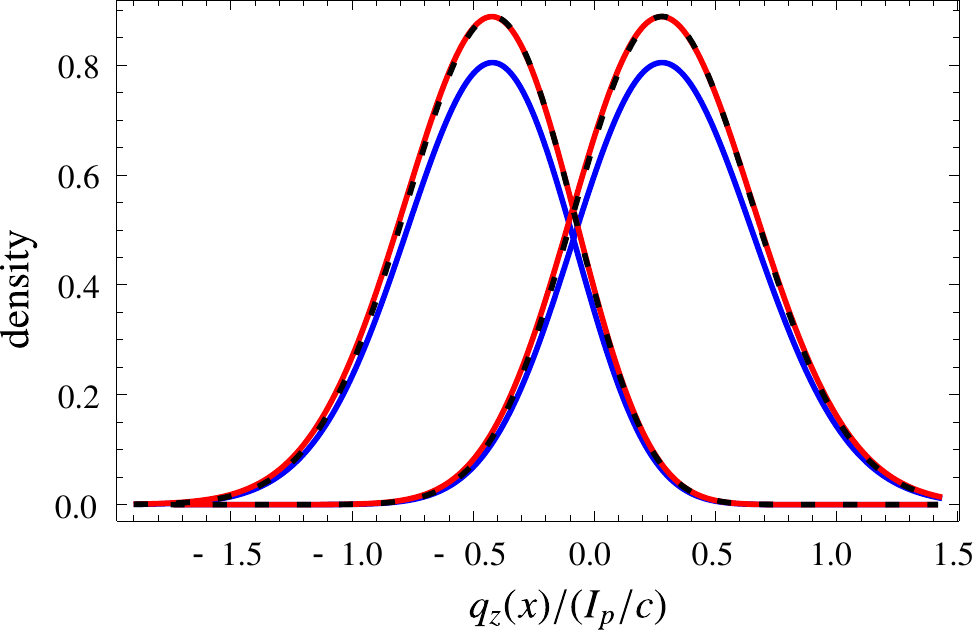}
  \caption{(Color online) Tunneling probability vs. the kinetic momentum
    along the laser's propagation direction at the tunnel entry (lines
    with peak on the left) and tunnel exit (lines with peak on the
    right).  Results for calculations including magnetic dipole
    correction are indicated in blue while for the red lines also the
    leading relativistic correction to the kinetic energy are taken into
    account. The dashed-black lines correspond to a fully relativistic
    calculation, which are very close to the ones including leading
    relativistic corrections to the kinetic energy.  The densities are
    normalized to the maximum density in the nonrelativistic case. A
    similar comparison was made in \cite{Klaiber_2013c} for the case of
    a zero-range potential via SFA.}
  \label{rel_mom}
\end{figure}

The energy level (\ref{epsilon_r}) depends on the $x$ coordinate.  This
is because the electron's kinetic momentum along the laser's propagation
direction $q_{z} (x) \equiv p_z-x E(t_0)/c$ changes during tunneling due
to the presence of the vector potential (magnetic field).  As a
consequence, the tunneling probability in the relativistic regime is
maximal at some non-zero canonical momentum $p_z$ in the laser's
propagation direction.  For instance, the kinetic momentum $q_z(x)$ with
maximal tunneling probability at the tunneling entry is
$q_{z}(x_0)\approx-0.42 I_p/c$, whereas at the exit it is $q_{z} (x_e)
\approx 0.28 I_p / c$ for the Coulomb potential, see
Fig.~\ref{rel_mom}.  During the under-the-barrier motion the
electron acquires a momentum kick into the laser's propagation direction
due to the Lorentz force, which can be estimated as
\begin{equation}
  \Delta p_z\sim x_e E_0/c\sim I_p/c,
  \label{Delta_p_z}
\end{equation}
with the barrier length $x_e\sim I_p/E_0$.  Thus, relativistic tunneling
can be visualized by an one-dimensional picture given in
Fig.~\ref{nrtp}(c). The area between the barrier and the position
dependent energy level is larger than in the nonrelativistic case due to
the non-vanishing transversal kinetic energy, indicating the reduced
tunneling probability in the relativistic description.

The WKB analysis can be carried out also in a fully relativistic way.
Taking into account the relativistic energy-momentum dispersion
relation, one obtains for the momentum and the ionization energy along
the polarization direction
\begin{equation}
  \varepsilon_x  =
  \sqrt{p_x^2 c^2 + c^4} + V_{\mathrm{barrier}}-c^2\,,
\end{equation}
\begin{equation}
  p_x (x) =
  \sqrt{\left(
      \frac{c^2-I_p -V_{\mathrm{barrier}}}{c} 
    \right)^2 -
    c^2 - p_y^2 - \left(
      p_z - \frac{x E(t_0)}{c} 
    \right)^2} 
\end{equation}
which determines the fully relativistic tunneling probability via
Eq.~(\ref{exponent}). The latter is shown in Fig.~\ref{rel_mom}. For
comparison it shows also the results for the nonrelativistic case, the
calculation using the magnetic dipole correction, and the calculation with
the leading relativistic kinetic energy correction $-\hat{p}_x^4/8c^2$
\footnote{The typical value of $p_z$ is already of order of $1/c$ and
  the $-\hat{p}_z^4/8c^2$ term is neglected as an higher order term in
  the $1/c$ expansion}.

As demonstrated in Fig.~\ref{rel_mom} the shift of the kinetic momentum
along the laser's propagation direction that maximizes the WKB tunneling
probability is determined mainly by the magnetic dipole correction to
the Hamiltonian.  This correction also decreases the tunneling
probability, since the 
Lorentz force due to the laser's transversal magnetic field transfers
energy from the tunneling direction into the perpendicular direction
hindering tunneling.  Taking into account further relativistic effects
does not change the behavior qualitatively but increases the tunneling
probability.  This can be understood intuitively by noticing that in the
reference frame of the relativistic electron the length of the barrier
is contracted and in this way enhancing the tunneling probability.  The
mass correction term is more important in the zero-range-potential case
than in the Coulomb-potential one, since the typical longitudinal
velocities are smaller in the latter case.  Furthermore,
Fig.~\ref{rel_mom} indicates that the calculation including only the
leading relativistic kinetic energy correction $\mathcal{O}(1/c^2)$
reproduces the fully-relativistic approach satisfactorily.  Thus, the
magnetic dipole and the leading order mass shift are the only relevant
relativistic corrections.

\begin{figure}
  \centering
  \includegraphics[scale=0.7]{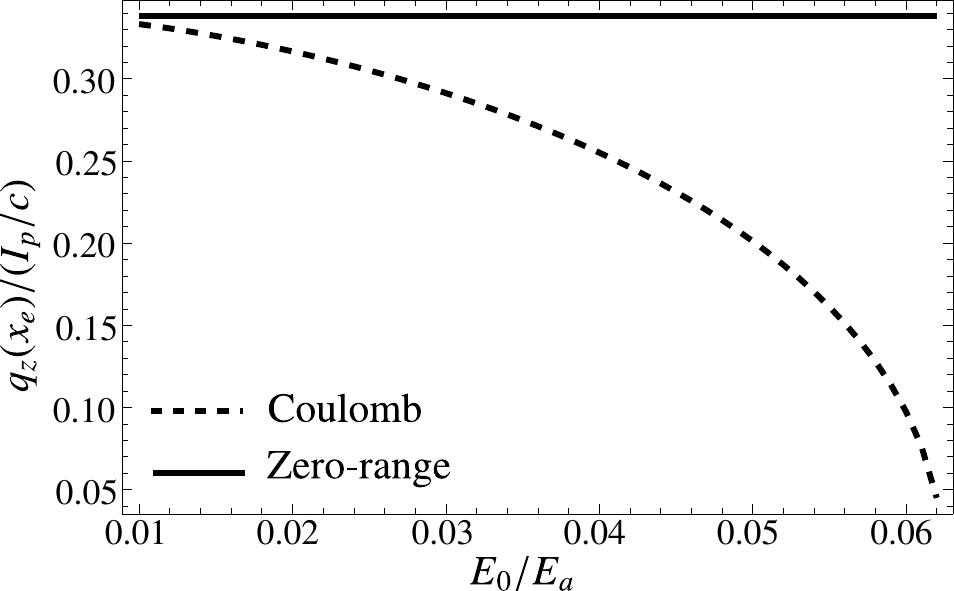}
  \caption{The kinetic momentum shift at the tunnel exit $q_{z}(x_e)$
    versus the barrier suppression parameter $ E_0/E_a $ for an electron
    bound by a Coulomb potential (dashed) and a zero-range atomic
    potential (solid).}
  \label{mom_shift_coulomb_vs_zero}
\end{figure}

The value of the kinetic momentum shift at tunnel exit $q_{z}(x_e)$
varies significantly with respect to the barrier suppression parameter
$E_0/E_a$ in the case of a Coulomb potential of the ionic core, as shown
in Fig.~\ref{mom_shift_coulomb_vs_zero}, while it does not depend on the
laser field in the case of zero-range atomic potential. The main reason
for the decreased momentum shift in the Coulomb potential case is that
the length of the Coulomb-potential barrier is reduced approximately by
a factor $(1-8{E_0}/{E_a})$ compared to the barrier length of the
zero-range potential.  According to Eq.~(\ref{Delta_p_z}), this barrier
length reduction leads to smaller momentum kick due to the magnetic
field.

\begin{figure}
  \centering
  \includegraphics[scale=0.525]{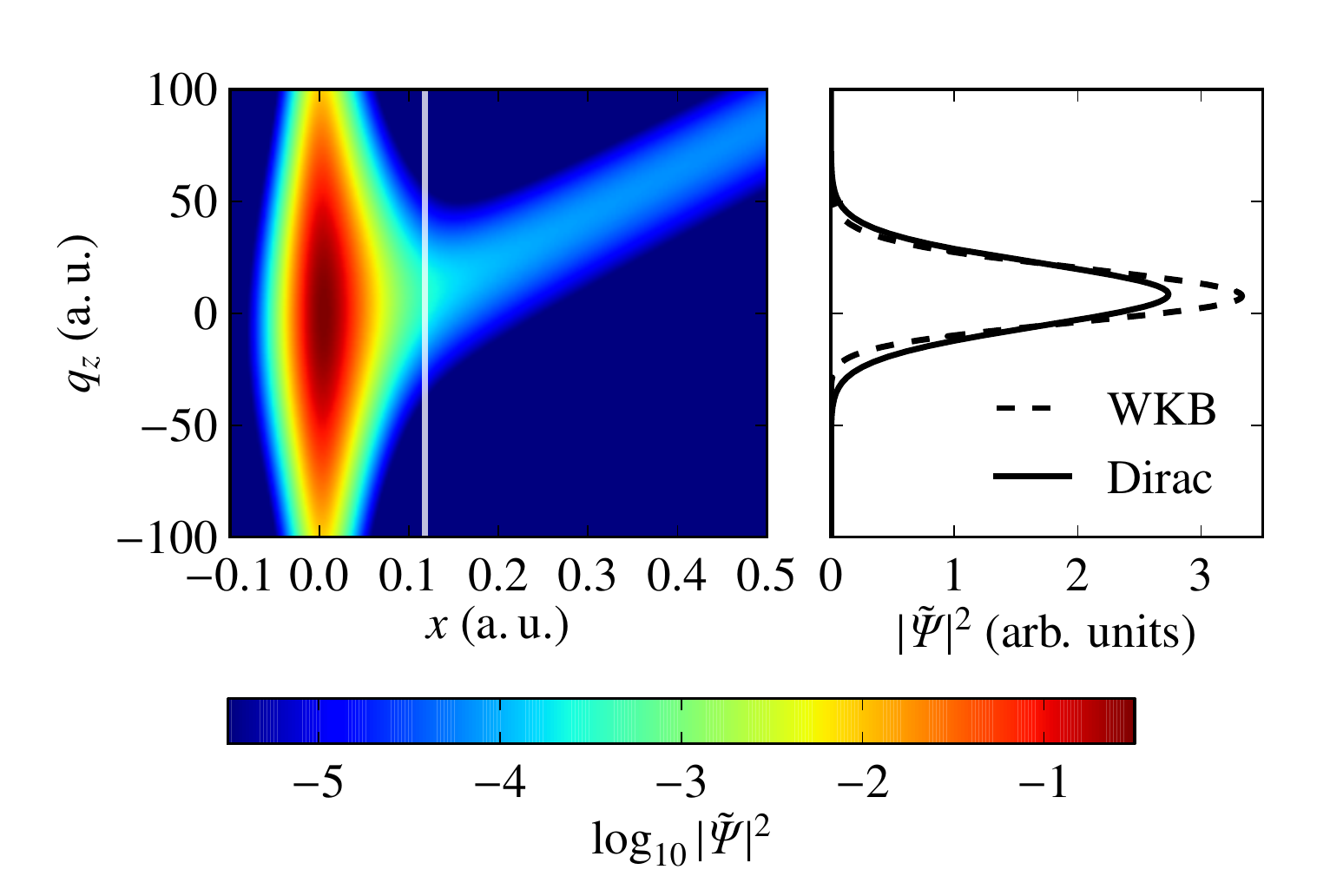}
  \caption{(Color online) Electronic density in the mixed space of
    position $x$ and kinetic momentum $q_z$ at the moment of maximal
    field strength at the atomic core (left panel). The electron's wave
    function has been obtained by simulating the tunnel-ionization from
    a two-dimensional soft-core potential by solving the time-dependent
    Dirac equation with all numerical parameters as in
    Fig.~\ref{fig:wf}.  The right panel shows the normalized kinetic
    momentum distribution of the tunneled electron at the tunnel exit
    (indicated by the white line in the left panel) as obtained by
    solving the Dirac equation and by the WKB approximation.}
  \label{fig:pos_kinmomentum_dist}
\end{figure}

Furthermore, we compare the prediction of the WKB approximation with the
results obtained by an \emph{ab initio} numerical calculation solving
the time-dependent Dirac equation \cite{Bauke_2011b}.  For this purpose
the tunnel-ionization from a two-dimensional soft-core potential was
simulated yielding the time-dependent real space wave function $\Psi(x,
z, t)$.  A transformation into a mixed representation of position $x$
and kinetic momentum $q_z$ via
\begin{equation}
  \tilde\Psi(x, q_{z}, t) = 
  \frac{1}{\sqrt{2\pi}}\int\Psi(x, z, t)
  e^{-i z(q_{z}-A_z(x, z))}\,d z
\end{equation}
allows us to determine the kinetic momentum in $z$ direction as a
function of the $x$ coordinate and in this way at the tunnel exit
$x=x_e$, see Fig.~\ref{fig:pos_kinmomentum_dist}.  Both, the solution of
the fully relativistic Dirac equation and the WKB approximation predict
a momentum distribution with a maximum shifted away from zero.  The
momentum shifts are in a good agreement.

\section{Tunnel-ionization with a zero-range potential model}
\label{sec:SFA}

The intuitive considerations of the previous section on the relativistic
under-the-barrier motion during tunnel-ionization led us to the
conclusion that relativistic tunneling induces a momentum kick along the
laser's propagation direction.  The aim of this section to prove this
conclusion by a rigorous calculation based on SFA, to show how this
momentum shift arises during the under-the-barrier motion, and to find
out how this relativistic signature is reflected in the electron
momentum distribution in far distance at the detector.

The SFA is based on an S-matrix formalism. The ionization is described by
the Hamiltonian
\begin{equation}
  H = H_0 + H_I (t)\,,
  \label{Hamiltonian}
\end{equation}
where $H_0$ is the field-free atomic Hamiltonian including the atomic
potential $V(\vec{x})$ and $H_I (t)$ denotes the Hamiltonian of the
laser-atom interaction. Initially at time $t \rightarrow - \infty$, the
electron is in the bound state $\ket{\psi(-\infty)} = \ket{\phi_0}$. In
SFA the influence of the atomic core potential on the free electron and the
influence of the laser field on the bound state are neglected.  This
allows us to express the time evolution of the state vector in the form
\cite{Becker_2002}
\begin{equation}
  \ket{\psi(t)} = 
  - i \int_{-\infty}^t dt' \, U_V (t,t') H_I (t') \ket{\phi_0 (t')}\,,
\end{equation}
where $U_V (t,t')$ is the Volkov propagator which satisfies 
\begin{equation}
 i \dfrac{\del U_V (t,t')}{\del t} =  H_V (t) U_V (t,t')
\end{equation}
with the Volkov Hamiltonian $H_V = H - V(\vec{x})$. 
The SFA wave function in momentum space of the final state reads
\begin{equation}
  \label{SFA_interaction}
  \braket{\vec{p}|\psi} = -i \int_{-\infty}^\infty dt' \, 
  \braket{\psi_V (t') | H_I  (t') | \phi_0 (t')}\,,
\end{equation}
where $\ket{\psi_V (t)}$ denotes a Volkov state \cite{Volkov_1935}. The
ionized part of the wave function in momentum space in
Eq.~(\ref{SFA_interaction}) can be expressed also in the form
\cite{Becker_2002}
\begin{equation}
  \label{SFA_coulomb}
  \braket{\vec{p}|\psi} = - i \int_{-\infty}^\infty dt' \, \bra{\psi_V (t')}
  V(\vec{x}) \ket{\phi_0 (t')}\,.
\end{equation} 
As the SFA neglects the effect of the atomic potential on the final
state, the SFA gives an accurate prediction when the atomic potential is
short ranged. For this reason, we will model the tunnel-ionization with
a zero-range potential in the following. The nonrelativistic tunneling
scheme for this case is visulized in Fig.~\ref{nrtp}(a). The barrier has
triangular shape which simplifies the analytical treatment of the
tunneling dynamics.

\subsection{Nonrelativistic case}

\begin{figure}
  \centering
  \includegraphics[width=\linewidth]{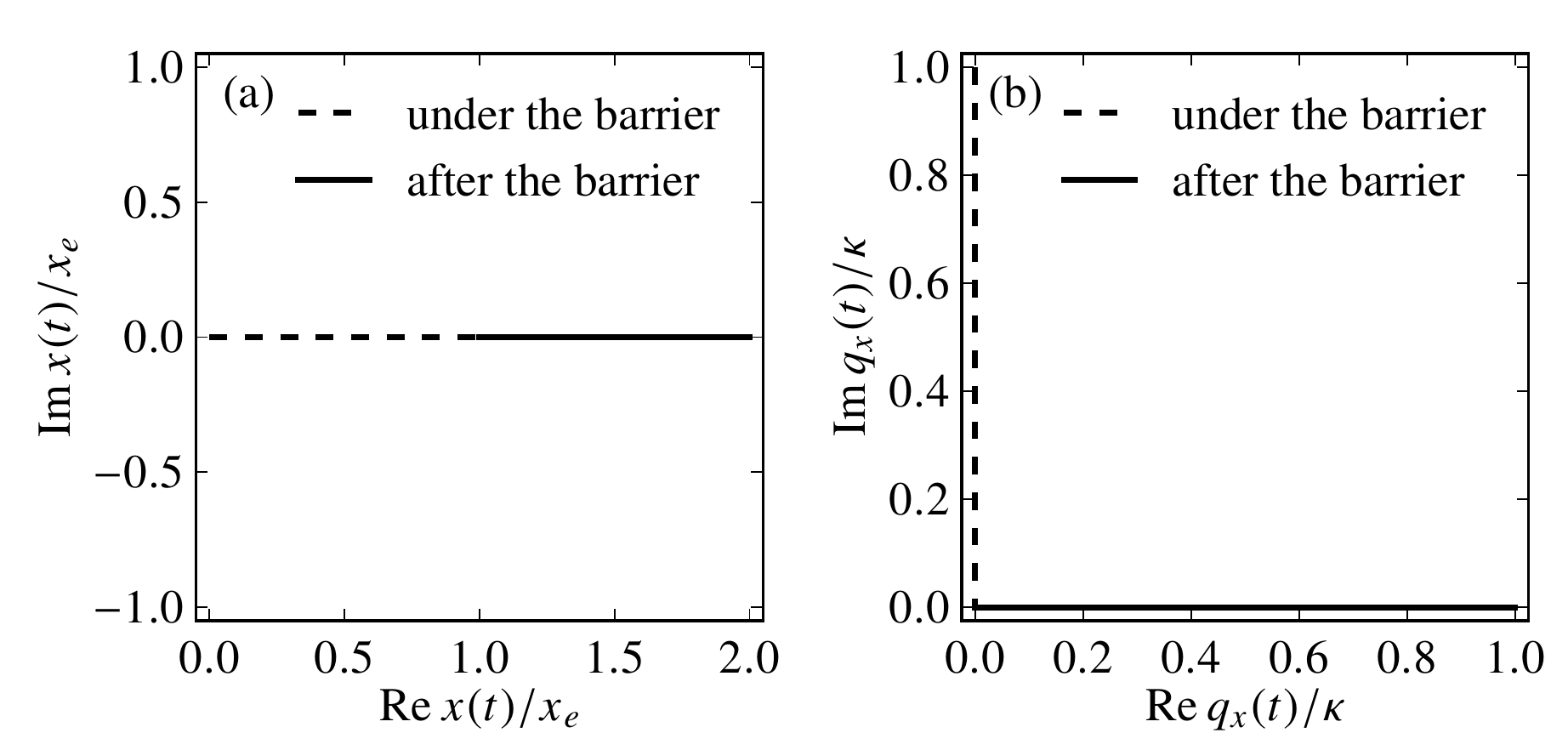}
  \caption{Coordinate (sub-figure (a)) and momentum (sub-figure (b))
    along the electric field direction of the complex SFA trajectory for
    tunnel-ionization from a zero-range potential.  During the
    under-the-barrier motion the momentum is imaginary whereas the
    coordinate is real.  Tunneling finishes when the kinetic momentum
    becomes real.}
  \label{non_rel_traject}
\end{figure}

Let us start our analysis with the nonrelativistic consideration when
the Hamiltonian for an atom in a laser field is given by
Eq.~(\ref{Hamiltonian}) with
\begin{align}
  H_0 &= \dfrac{p^2}{2} + V^{(0)}(\vec{x})\,, \\
  H_I &= \vec{x} \cdot \vec{E}(t)\,,
\end{align} 
where $V^{(0)}(\vec{x})$ is the zero-range atomic potential. In SFA the
ionized part of the wave function far away after the laser field has
been turned off reads \cite{Klaiber_2013a}
\begin{equation}
  \label{non_rel_mom_wave}
  \braket{\vec{p}|\psi}  = 
  - i \mathcal{N} \int_{-\infty}^\infty dt \, e^{-i \tilde{S}(\vec{p},t)} \,,
\end{equation}
where 
\begin{equation}
\label{contracted_action_nonrel}
\tilde{S}(\vec{p},t) = -\kappa^2 t/2-\int^{t} dt' \, \vec{q}^2/2
\end{equation}
is the contracted action, $\vec{q} = \vec{p}+\vec{A}/c $ is the kinetic
momentum, $\vec{A} = - c\int^t \vec{E} \,dt'$,
$\mathcal{N}\equiv\bra{\vec{q}}V^{(0)} \ket{{\phi^{(0)}}}=\rm const$,
and $\ket{{\phi^{(0)}}} e^ {i \kappa^2 t / 2}$ is the bound state of the
zero-range potential.  The time integral in Eq.~(\ref{non_rel_mom_wave})
can be calculated via the saddle point approximation (SPA). The saddle
point equation
\begin{equation}
  \dot{\tilde{S}}(\vec{p},t_s) = q(t_s)^2 + \kappa^2= 0
\end{equation}
yields the kinetic momentum $q(t_s) = i \kappa$ at the saddle point time
$t_s$. Then, the wave function in momentum space reads in the quasi-static limit
\begin{equation}
  \label{eq:psi_p_nr}
  \braket{\vec{p}|\psi} = 
  -i \mathcal{N} \sqrt{\dfrac{2 \pi }{|E(t_s)|\sqrt{p_\perp^2+\kappa^2}}} 
  \exp\left[-\frac{(p_\perp^2 + \kappa^2)^{3/2}}{3 |E(t_s)|}\right]
\end{equation}
for the vector potential $\vec{A} ={c E_0} \sin(\omega t)/{\omega}
\hat{\vec{x}}$ and with $|E(t_s)| = E_0 \sqrt{1-(p_x/(E_0 /\omega))^2} $
and $p_\perp = \sqrt{p_y^2 + p_z^2}$.  From expression
\eqref{eq:psi_p_nr} it follows that the density of the ionized wave
function is maximal at $p_\perp = 0$ for any value of $p_x$.  The
coordinate space wave function 
\begin{equation}
  \braket{\vec{x}|\psi} =  - i \dfrac{\mathcal{N}}{(2 \pi)^{3/2}}
  \int_{-\infty}^\infty dt  \int d^3 p \, \exp\left[i \vec{x} \cdot \vec{p} -i
    \tilde{S}(\vec{p},t)\right]
  \label{int}
\end{equation}
is obtained by a Fourier transform of Eq.~(\ref{non_rel_mom_wave}).
From the SPA it follows that the main contribution to the integral over
$\vec{p}$ in Eq.~(\ref{int}) originates from momenta near the momentum
which fulfills the saddle point condition \mbox{$\vec{x} -
  \del_{\vec{p}}\tilde{S}(\vec{p},t)=0$}. The latter defines the
trajectories 
$\vec{x}=\vec{x}(\vec{p},t)$ which contribute to the transition
probability with amplitudes depending on $\vec{p}$. For the most
probable final momentum $\vec{p}_0 = 0$, the trajectory which start at
the tunneling entry at time $t_s$ is given by
\begin{equation}
  \label{eq:x_t_nr}
  \vec{x}(t) = \del_{\vec{p}} \tilde{S}(\vec{p}_0,t) = 
  \int_{t_s}^t \, dt' \dfrac{\vec{A}(t')}{c} \,.
\end{equation}
The line integral in Eq.~\eqref{eq:x_t_nr} is along a path connecting
the complex time $t_s$ 
with the real time $t$.  The complex saddle point time $t_s$ can be
determined by solving $q(t_s) = i \kappa$ for $t_s$. The corresponding
kinetic momentum is
\begin{equation}
  \label{eq:p_t_nr}
  \vec{q}(t) = \dfrac{\vec{A}(t)}{c}\,.
\end{equation}
In Fig.~\ref{non_rel_traject} the complex trajectory \eqref{eq:x_t_nr}
and the complex kinetic momentum \eqref{eq:p_t_nr} along the tunneling
direction are shown. The spatial coordinate is real under the barrier as
well as behind the barrier, whereas the kinetic momentum is imaginary
during tunneling and becomes real when leaving the barrier, which
corresponds to the time $\RE[t_s]$.  The tunneling exit coordinate is
$x_e=x_e(\RE[t_s])=I_p/E_0$ which is consistent with the intuitive
tunneling picture.  The momentum in the tunneling direction is $q(t_s) =
i\kappa$ when tunneling starts, whereas it is $q(\RE[t_s])=0$ at the
tunnel exit.

\subsection{Relativistic case}

Our fully relativistic consideration is based on the Dirac Hamiltonian
with a zero-range atomic potential
\begin{equation}
  H = c \vec{\alpha} \cdot (\vec{p} + \vec{A}/c) 
  - \phi + \beta c^2  + V^{(0)} (\vec{x})
\end{equation}
where $\vec{\alpha}$ and $\beta$ are standard Dirac matrices
\cite{Landau_4} and the G{\"o}ppert-Mayer gauge~(\ref{GM}) is employed. The
ionized part of the momentum wave function in SFA yields
\begin{equation}
  \braket{\vec{p}|\psi} = 
  -i \int_{-\infty}^\infty dt \int d^3 x \,
  \overline{\psi}_V(\vec{x},t) \gamma^0 V^{(0)} (\vec{x}) \, \phi^{(0)}
(\vec{x},t)
  \,.
\end{equation}
Here $\phi^{(0)} (\vec{x},t)= e^{-i \varepsilon_0 t} \varphi^{(0)}
(\vec{x}) {v^{(0)}}_\pm$ is the ground state of the zero-range potential
with the ground state spinor ${v^{(0)}}_\pm$ and $\varepsilon_0 =c^2
-I_p $; $\psi_V (\vec{x},t) = N_V u_{\pm}e^{i S}$ is the relativistic
Volkov wave function in the G{\"o}ppert-Mayer gauge for a free electron in a
laser field, which is obtained from the Volkov \textbf{wave} function in the 
velocity
gauge \cite{Volkov_1935} with further gauge transformation via the gauge
function $\chi = - \vec{x}\cdot \vec{A}/c$, $\vec{A}(\eta) \equiv -
\dfrac{c}{\omega}\int^\eta \vec{E}(\eta') d\eta'$.  Furthermore, $N_V$
is the normalization constant,
\begin{equation}
  \label{eq:S_rel}
  S = - \varepsilon t + \left(\vec{p}+\frac{\vec{A}}{c}\right)\cdot \vec{x} -
  \dfrac{1}{c
    \Lambda}
  \int^\eta \left(\vec{p}\cdot \vec{A}+\frac{\vec{A}^2}{2c}\right)d\eta'\,,
\end{equation}
is the quasiclassical action and 
\begin{equation}
  u_{\pm} = \left(1+ \dfrac{\omega}{2 c^2
      \Lambda}(1+\vec{\alpha}\cdot \hat{\vec{k}})\vec{\alpha} \cdot
\vec{A}\right)
  {u_0}_\pm,
\end{equation}
with $\Lambda = p^\mu k_\mu = \omega (\varepsilon / c^2 - \vec{p} \cdot
\hat{\vec{k}}/c)$, $p^\mu = (\varepsilon/c, \vec{p})$, and the free
particle spinor ${u_0}_\pm$.  After averaging over the spin of the
initial electron as well as over the spin of the ionized electron the
wave function of the ionized electron reads
\begin{equation}
  \label{rel_mom_space_wave_function}
  \braket{\vec{p}|\psi} = - i \dfrac{N_V (2
    \pi)^{3/2}}{2\omega}\int_{-\infty}^\infty d \eta \, e^{-i \tilde{S} }
  \sum_{s,s'}  \bra{\vec{q_d},s'} V^{(0)} (\vec{x}) \ket{\varphi_0, s}
  \,.
\end{equation}
Here, a coordinate transformation $(t,\vec{x})\rightarrow(\eta,\vec{x})$
is employed and
\begin{equation}
\label{contracted_action_rel}
  \tilde{S} = \dfrac{1}{2 \Lambda} \left( -\kappa^2 \eta -  \int^\eta
\vec{q_d}^2 d\eta' \right)
\end{equation}
is the contracted action with the field-dressed electron momentum in
the laser field
\begin{equation}
  \vec{q_d} = 
  \vec{p} + \frac{\vec{A}}{c} - 
  \frac{\hat{\vec{k}}(\varepsilon-\varepsilon_0)}{c}\,.
\end{equation}
For a zero-range potential the inner product $\bra{\vec{q_d},s'} V^{(0)}
(\vec{x}) \ket{\varphi_0, s}$ is only $\eta$ dependent. Further, the
$\eta$-integral in \eqref{eq:S_rel} can be calculated using SPA and the
saddle point equation yields $\vec{q_d}^2 (\eta_s) = - \kappa^2 $ as in
the nonrelativistic regime \cite{Klaiber_2013b}. Then, the wave function
of the ionized electron in momentum space yields 
\begin{equation}
  \label{final_wave_fucntion_mom_space}
  \braket{\vec{p}|\psi} = 
  -i \mathcal{N}(\eta_s) \sqrt{\dfrac{2 \pi
      \Lambda}{\omega |E(\eta_s)|\sqrt{{q_d}_\perp^2 +\kappa^2}}}
  \exp\left[ - \dfrac{\omega({q_d}_\perp^2 +\kappa^2)^{3/2}}{3 \Lambda 
|E(\eta_s)|}
  \right]
\end{equation}
for the vector potential $\vec{A} = c E_0 \sin(\eta)/\omega
\hat{\vec{x}}$ with the laser's propagation vector
$\hat{\vec{k}}=\hat{\vec{z}}$, where
\begin{align}
  |E(\eta_s)| &= E_0 \sqrt{1-(p_x/(E_0 /\omega))^2} \,,\\
  {q_d}_\perp &= \sqrt{p_y^2 + \left(p_z
      -\frac{\varepsilon-\varepsilon_0}{c}\right)^2}\,, \\
  \mathcal{N}(\eta_s) &= \dfrac{N_V (2\pi)^{3/2}}{2 \omega} \sum_{s,s'}
  \bra{\vec{q_d},s'} V^{(0)} (\vec{x}) \ket{\varphi_0, s}\,.
\end{align}

\begin{figure}
  \centering
  \includegraphics[scale=0.6]{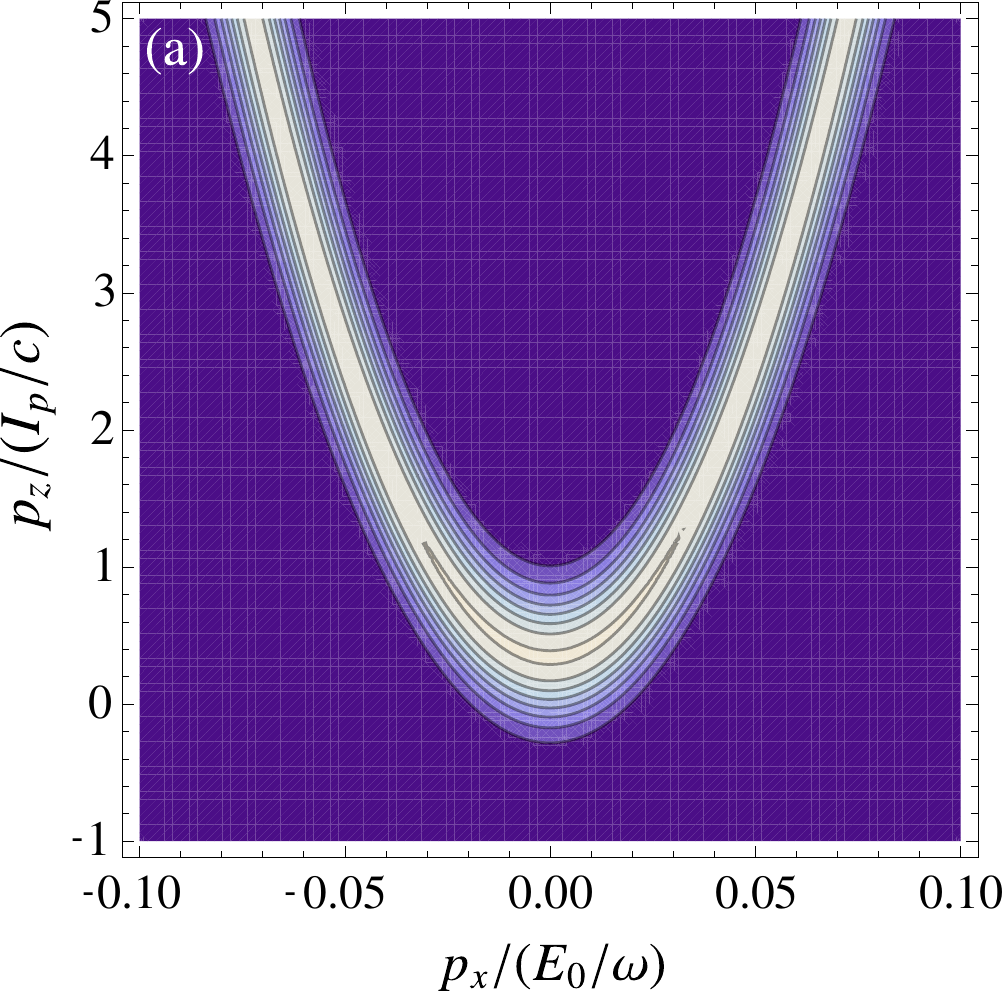}\\[3ex]
  \includegraphics[scale=0.6]{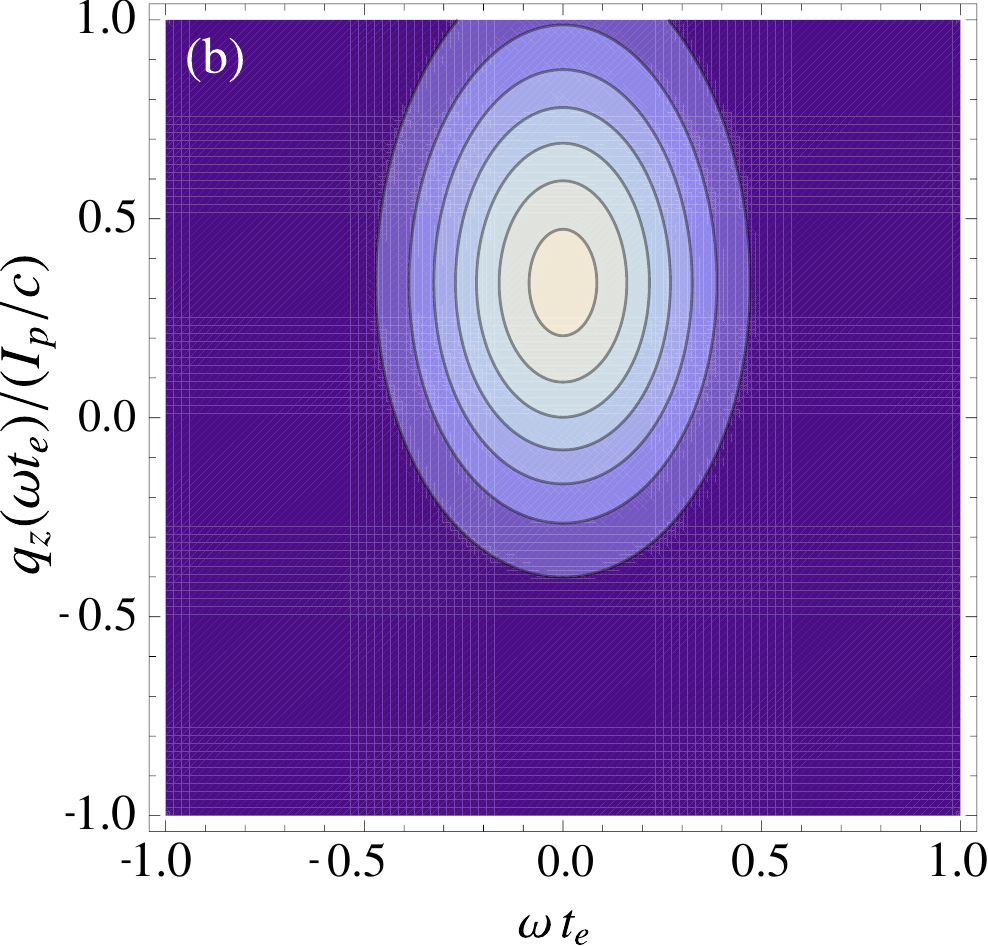}
  \caption{(Color online) The momentum distribution of the ionized
    electron: (a) in infinity at the detector; (b) at the tunnel exit,
    depending on the tunnel exit time $t_e$.  The maximum tunneling
    probability occurs at $p_z = I_p /(3 c)$. The applied parameters are
    $\kappa = 90$, $E_0/E_a = 1/30$ and $\omega = 10$.}
\label{rel_momentum_shift}
\end{figure} 

The relativistic momentum distribution of the ionized electron of
Eq.~\eqref{final_wave_fucntion_mom_space} differs qualitatively from the
nonrelativistic one.  In the nonrelativistic case the maximum of the
distribution is at $p_{\perp}=0$ at any $p_x$. In the relativistic case,
however, the momentum distribution has a local maxima along the parabola
which can be approximated as
\begin{equation}
  \label{rel_most_probable_momentum}
  p_z \approx 
  \dfrac{I_p}{3 c}\left(1+ \dfrac{I_p}{18 c^2} \right) +
  \dfrac{p_x^2}{2c}\left(1+\dfrac{I_p}{3c^2} + \dfrac{2 I_p^2}{27 c^4}\right) +
  O\left(\frac{I_p^2}{c^4}\right)
  \,,
\end{equation}
see Fig.~\ref{rel_momentum_shift}(a).  The global maximum of the
tunneling probability is located at $p_z = I_p /(3 c)$, while in the
nonrelativistic case it is at $p_z =0$.  This shift of the maximum is
connected with the first step of the ionization, the tunneling, whereas
the parabolic wings are shaped in the second step, the continuum
dynamics.  These wings are located around $p_z=U_p/c$ with the
ponderomotive potential $U_p = E_0^2/(4 \omega^2)$.

The momentum distribution at the tunnel exit can be calculated via back
propagation of the final momentum space wave function
(\ref{final_wave_fucntion_mom_space}). Thus, the wave function at the
tunnel exit is\pagebreak[1]
\begin{multline}
  \braket{\vec{p}|\psi(t_e)} = \int d^3 p' \bra{\vec{p}} U(t_e, t_f)
  \ket{\vec{p'}}\braket{\vec{p'}|\psi(t_f)} \\
  \approx \int d^3 p' \bra{\vec{p}} U_V (t_e, t_f)
  \ket{\vec{p'}}\braket{\vec{p'}|\psi(t_f)}
\end{multline}
with the tunnel exit time $t_e=\RE[t_s]$ and final time $t_f$ where the
interaction is turned off, hence the Volkov wave function reduces to the
free particle wave function.  Because $\omega\hat{\vec{k}}\cdot \vec{x}
\ll c $ holds at the tunnel exit holds, the exact Volkov propagator
\begin{equation}
  \bra{\vec{x}}U_V (t,t')\ket{\vec{x'}} = 
  \int d^3 p \, \, \psi_V (t,\vec{x})
  \psi_V^{\dagger} (t',\vec{x'})
\end{equation}
can be simplified expanding the phase dependent functions around $\omega
t_e$, which yields
\begin{equation}
  U_V (t_e,t_f) = \int d^3 p \, \exp\left(i \varphi(t_e,t_f)\right)
  \ket{\vec{p_e}}\bra{\vec{p}}
\end{equation}
with the exit momentum and the phase
\begin{align}
  \vec{p_e} &= \vec{p} +\dfrac{\vec{A}(\omega t_e)}{c} + \hat{\vec{k}}
  \dfrac{\omega}{c^2 \Lambda}
  (\vec{p}+\dfrac{\vec{A}(\omega t_e)}{2 c})\cdot \vec{A}(\omega t_e)\,, \\
  \varphi(t_e,t_f) &= \varepsilon (t_f - t_e ) + \dfrac{1}{c \Lambda}
  \int^{\omega
t_e} d\eta \,  \left(\vec{p}+ \dfrac{\vec{A}(\eta)}{2c}\right) \cdot  \vec{A}(\eta)\,,
\end{align}
respectively. As a result, the momentum space wave function at the
tunnel exit $t_e$ reads in terms of the final wave function
$\braket{\vec{p'}|\psi(t_f)}$
\begin{equation}
  \braket{\vec{p}|\psi(t_e)} = e^{i \varphi(t_e,t_f)}
  \braket{\vec{p'}|\psi(t_f)}
\end{equation}
with 
\begin{equation}
  \vec p' = \left( 
     -\dfrac{A_x (\omega t_e)}{c}, 
    p_y,
    p_z+\frac{A_x (\omega t_e)^2 \omega }{2 c^3 \Lambda }
  \right)\,.
  \label{mom_at_exit}
\end{equation}
The transversal momentum distribution at the tunnel exit can be
calculated via replacing the momentum in the wave function
Eq.~(\ref{final_wave_fucntion_mom_space}) with Eq.~(\ref{mom_at_exit}),
which can be seen in Fig.~\ref{rel_momentum_shift}(b). The comparison of
Figs.~\ref{rel_momentum_shift}(a) and~(b) indicates that the
relativistic shift of the peak of the transverse momentum distribution
at the tunnel exit $p_z=I_p/(3c)$ is maintained in the final momentum
distribution. The parabola can for example be calculated from classical
trajectories.  The kinetic momentum at the exit is connected with the
final momenta via
\begin{align}
  \nonumber
  q_x(\eta_s)&= p_x+\frac{A(\eta_s)}{c}=0\,,\\
  q_z(\eta_s)&= p_z+\frac{\omega}{c^2 \Lambda} \left( p_xA(\eta_s)+\frac{A(\eta_s)^2}{2 c}
\right) =\frac{I_p}{3c}
\end{align}
and the relation 
\begin{eqnarray}
  p_z=\frac{I_p}{3c}+\frac{\omega p_x^2}{2c \Lambda}
\end{eqnarray}
follows.

\begin{figure}
  \centering
  \includegraphics[scale=0.6]{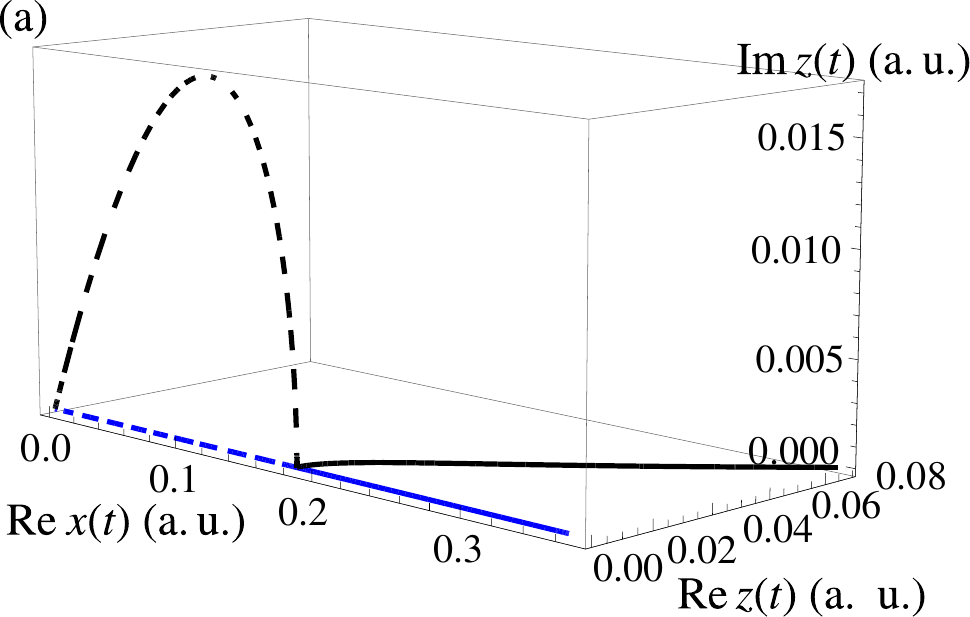}\\[2ex]
  \includegraphics[scale=0.6]{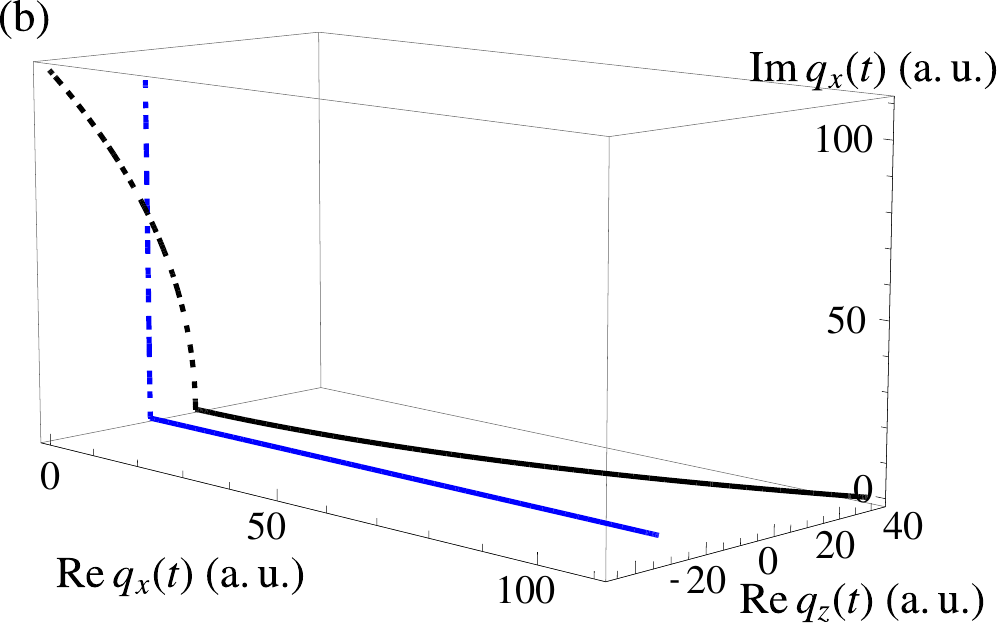}
  \caption{(Color online) Comparison of the nonrelativistic (blue) and
    relativistic (black) complex SFA trajectories: (a) coordinates and
    (b) momentum components. The dashed and the solid lines correspond
    to the under-the-barrier motion and the motion after the tunneling,
    respectively. In the relativistic regime, the trajectory enters the
    tunneling barrier at $(0,0)$, it is complex under the barrier and
    becomes real again when it leaves the tunneling barrier. The
    trajectory enters the barrier with the transversal momentum $-2 I_p
    /(3 c)$ and leaves the barrier with $I_p/(3 c)$. The applied
    parameters are $\kappa = 90$, $E_0/E_a = 1/30$.}
  \label{rel_traject}
\end{figure}

We investigate the trajectory of the electron and its momentum during
the tunneling in the relativistic regime. The coordinate wave function
can be obtained via a Fourier transform of momentum space wave
function~(\ref{rel_mom_space_wave_function}). Then, the stationary phase
condition gives the quasiclassical trajectories at the most probable
momentum given by Eq.~(\ref{rel_most_probable_momentum}). The results
are plotted in Fig.~\ref{rel_traject}.  It shows that in the
relativistic regime most probable trajectory is the trajectory where the
electron enters the barrier with the transversal momentum $-2 I_p /(3
c)$ and reaches the exit with $I_p/(3 c)$.  This is in accordance with
our intuitive discussion in Sec.~\ref{sec:intuitive}.

For the most probable momentum, the trajectory starts at the real axis,
obtains complex values during tunneling and has to return to the real
axis after tunneling Fig.~\ref{rel_traject}(a).  Here an analogy to
relativistic high-harmonic generation can be drawn, where also a return
condition has to be fulfilled: the recollision of the ionized electron
to the atomic core in the presence of the drift motion induced by the
laser's magnetic field \cite{Klaiber_2007}. Similarly, the electron has
to start with a momentum of the order of $ U_p / c$ against the laser
propagation direction (cf.\ $q_{z,\rm i}=-2 I_p /(3 c)$ at the entering
the ionization barrier) which compensates the drift motion and
facilitates the recollision to the atomic core with a momentum $U_p / c$
along the laser's propagation direction (cf.\ $q_{z,\rm e}= I_p /(3c)$ at
the exit of the ionization barrier).

The shift of the electron's momentum distribution along the laser's
propagation direction in the relativistic regime is possible to detect
by measuring the final momentum distribution of the ion
\cite{Smeenk_2011}.  The ionized electron acquires momentum along the
laser's propagation direction in the laser field because of the absorbed
momentum of the laser photons. However, part of the momentum of laser
photons is transferred to the ion. The energy conservation law provides
a relationship between the number of absorbed photons $n$ and the
electron momentum $p_e$
\begin{equation}
  n\omega-I_p+c^2\approx\varepsilon_e\,,
\end{equation}
where $\omega$ is the laser frequency, $I_p$ the ionization potential,
and $\varepsilon_e=c\sqrt{p_e^2+c^2}$ the energy of the electron. The
kinetic energy of the ionic core can be neglected due to the large mass
of the ion.  Additionally, the momentum conservation law gives
information on the sharing of the absorbed photon momentum between the
ion and the photoelectron
\begin{equation}
  \begin{pmatrix}
    {n\omega}/{c} \\
    0 \\
    0 
  \end{pmatrix} =
  \begin{pmatrix}
    p_{e\,z}+p_{0\,z}\\
    p_{e\,x}+p_{0\,x} \\
    p_{e\,y}+p_{0\,y}
  \end{pmatrix}\,.
\end{equation}
In the nonrelativistic tunneling regime of ionization in the linearly
polarized laser field, the photoelectron's most probable momentum is
$p_{e}=0$, when $n\omega=I_p$ and the ion carries out a momentum
$p_{0\,z}=I_p/c$. In the relativistic regime of interaction the most
probable value of the photoelectron's momentum is not vanishing but
equals $p_{e\,z}\approx I_p/3c$ in the case of linear polarization. In
this case the momentum conservation will provide the ion momentum
$p_{0\,z}\approx 2I_p/3c$. In \cite{Smeenk_2011,Titi_2012} the momentum
sharing between ion and electron during the tunnel-ionization in a
strong circularly polarized laser field is investigated. Their result
supports the simple-man model prediction. It is shown that the total
momentum of the absorbed photons, that is $I_p/c$, is transferred to the
ion and not to the ionized electron. The difference with respect to our
result may be explained by the focal averaging as well as by the fact
that our parameters are situated in the pure tunneling regime whereas
the references deal with ionization at the transition to
over-the-barrier ionization. The momentum shift of the ionized electrons
at the detector that we describe is a genuine feature of the
relativistic tunneling dynamics.

\subsection{Tunneling formation time}
\label{formation_time_s}

\begin{figure}
  \centering
  \includegraphics[scale=0.7]{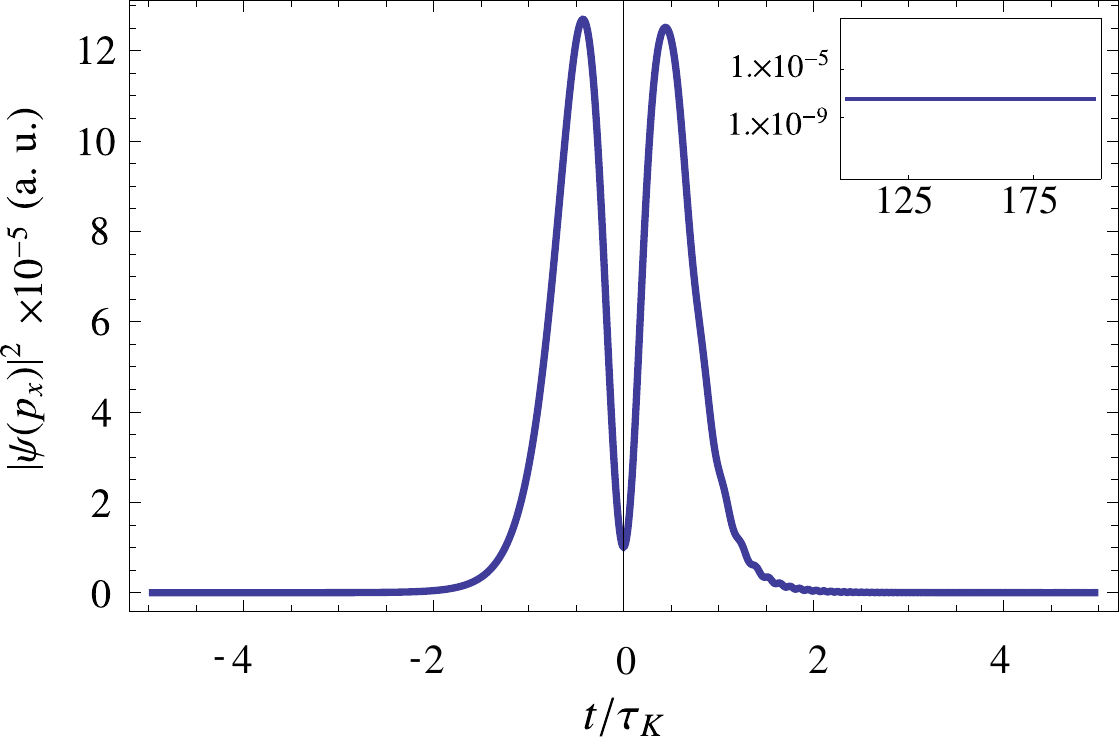}
  \caption{The ionization momentum amplitude for $p_x=0$ vs.\ the
    observation time $t$. The inset shows the momentum amplitude for
    large times, which coincides with the ionization amplitude.  The
    applied parameters are $E_0/E_a = 1/30$ and $\kappa = 1$.}
  \label{formation_time}
\end{figure}

The physical interpretation of the Keldysh time as the ionization
formation time given in Sec.~III B can be readily clarified
within the SFA formalism.  In the SFA, the momentum wave function in the case of a
zero-range atomic potential at some intermediate time is given by
\begin{equation}
  \langle p_x|\psi (t) \rangle = -i \int_{-\infty}^t dt' \exp[-i \tilde{S}(t,t')]
\braket{\vec{q}(t')|H_I (t')|\phi_0} \,,
  \label{formation}
\end{equation}
where $\braket{\vec{q}(t')|H_I (t')|\phi_0}$ is the pre-exponential factor that slowly varies with
time and the quasiclassical action $\tilde{S}(t,t')$ is given either by
Eq.~(\ref{contracted_action_nonrel}) for the nonrelativistic case, or by
Eq.~(\ref{contracted_action_rel}) for the relativistic case.

A numerical integration of Eq.~(\ref{formation}) for a monochromatic laser field in the
nonrelativistic regime is shown in Fig.~\ref{formation_time}. The value of the momentum amplitude
starts with zero at early times and then varies on a time scale that is of the order of the
Keldysh time.  Thus, the Keldysh time is the typical time scale for the formation of the moment
components of the ionized wave function, or in short, the formation time of the
ionization.  For large times, the momentum amplitude stabilizes on some
positive value, which can be identified as the ionization amplitude of
the specific momentum component, see inset in Fig.~\ref{formation_time}.
We note that the qualitative behavior of the time evolution of the 
amplitude shown in Fig.~\ref{formation_time} remains the same for the nonrelativistic as 
well as for the relativistic ionization from a Coulomb potential.


\section{General aspects of tunneling time delay}
\label{sec:tunneling_time}

In quantum mechanics time as it stands is a parameter and not an
observable. Furthermore, as it is discussed by Pauli \cite{*[{}]
  [{\space p.~60.}] Pauli_1926,Carruthers_1968}, it can not be upgraded
to an operator which is conjugate to the Hamiltonian. It is due to the
fact that although the time can take any values, the spectrum of allowed
energy levels of a given Hamiltonian can not span the entire real
line. Namely, either the Hamiltonian is bounded below or it may take
discrete values due to the quantization conditions. Nevertheless, a time
delay problem can be formulated in quantum mechanics. It is possible to
put forward in a reasonable way within quantum mechanics the question on
how much time an electron spends in a specified space region during its
motion and, in particular, what is the time delay for the tunneling
through a potential barrier
\cite{MacColl_1932,Eisenbud_1948,Wigner_1955,Smith_1960,Baz_1966,Galapon_2012,Sokolovski_2008,
Peres_1980,Landauer_1994,Davies_2005}.
Several definitions for the time delay have been proposed, one of the
most accepted definitions is the Eisenbud-Wigner-Smith time delay (here
refered to as the Wigner time delay.)  \cite{Eisenbud_1948, Wigner_1955,
  Smith_1960}.

The definition of the Wigner time delay is based on the trajectory of the peak of the electron wave
packet, as long as the wave packet has a unique peak. Let us illustrate it considering the motion of
the following wave packet in the position space
\begin{equation}
  \braket{x|\psi(t)} = 
  \dfrac{1}{\sqrt{2 \pi}} \int_{-\infty}^{\infty} d p \,
  \exp\left(i p x\right) \braket{p|\psi(t)}\,,
\end{equation}
assuming that the wave packet in momentum space $\braket{p|\psi(t)}$ is
centered around $p_0$ and is expressed in the form
\begin{equation}
  \braket{p|\psi(t)} = g(p)\exp\left(-i \phi(p,t)\right),
\end{equation}
with real functions $g$ and $\phi$. The peak of the wave packet in
position space at a moment $t$
\begin{equation}
  \braket{x|\psi(t)} = 
  \dfrac{1}{\sqrt{2 \pi}} \int_{-\infty}^{\infty} d p
  \, \exp\left(i \left(p x - \phi(p,t)\right)\right) g(p)
\end{equation}
can be found in the limit $\Delta p_g \gg \phi'(p_0)$, with the width
$\Delta p_g$ of the density $g$, by the stationary phase approximation,
and is given by the stationary phase condition
\begin{equation}
  \label{expectation_from_phase} 
  \left. 
    \dfrac{\del}{\del p} (p x - \phi(p,t))
  \right|_{p_0} = 0 
  \quad \Rightarrow \quad 
  x = \left. \dfrac{\del \phi(p,t) }{\del p}\right|_{p_0}\,.
\end{equation}
In the limit $\Delta p_g \ll \phi'(p_0)$, the phase $\phi$ can be
linearized, viz., $\phi(p)=\phi(p_0) + \phi'(p_0) (p - p_0 ) $ and the
maximum of the wave-packet is shifted from $ 0 $ to $ \phi'(p_0) $ due
to the additional coordinate translation operator $\exp\left[ i p
  \phi'(p_0)\right]$. Therefore, in both cases the phase derivative at
$p_0$ yields the coordinate of the maximum of the wave-packet. In fact,
this result is consistent with the expectation value of the position
operator
\begin{equation}
  \langle x \rangle = \bra{\psi(t)} x \ket{\psi(t)} =
  \int_{-\infty}^{\infty} d
  p\, \left( i g(p) g(p)' + g(p)^2
    \phi(p,t)'\right)\,.
\end{equation}
In the latter, the first term vanishes because the wave packet is
initially formed symmetrically around $p_0$, while from the second
term one has $\langle x \rangle \approx \phi(p_0,t)'$, if the phase
$\phi(p,t)'$ is expanded around $p_0$ and the third and higher-order
derivatives are neglected.

Similarly, the wave packet in position space can be expanded in energy
eigenfunctions with energy eigenvalues $\varepsilon$ \cite{*[{}] [{\space
    p.~31.}] merzbacher_1961}
\begin{multline}
  \braket{x|\psi(t)} = \int_{0}^{\infty} d \varepsilon \,
  \braket{x|\varepsilon}
  \braket{\varepsilon|\psi(t)} \\
  = \int_{0}^{\infty} d\varepsilon \, f(x,\varepsilon) \exp\left(i
    \varphi(x,\varepsilon)-i \varepsilon
    (t-t_0)\right)\,,
\end{multline}
where $f(x,\varepsilon)\equiv \braket{\varepsilon|\psi(t_0)}
|\braket{x|\varepsilon}| $ and $\braket{\varepsilon|\psi(t_0)}$ is the
energy distribution of initial wave packet at $t=t_0$ which is
symmetrically centered around $\varepsilon_0$, and
$\varphi(x,\varepsilon)$ is the phase of $\braket{x|\varepsilon}$ which
is the steady-state solution. Analogously to the dicussion on the
coordinate maximum, the condition
\begin{equation}
  \label{definition_of_time}
  \tau \equiv t-t_0 = 
  \left. 
    \frac{\del \varphi(x,\varepsilon) }{\del\varepsilon}
  \right|_{\varepsilon_0},
\end{equation}
then determines the moment when the wave packet is maximal at a given
point with coordinate $x$, i.\, e. , the Wigner trajectory. Equation~\eqref{definition_of_time}
indicates that the phase of the steady state solution to the Schr{\"o}dinger
equation is sufficient to deduce the Wigner trajectory.  The difference
between the Wigner trajectory and the classical trajectory (no time interval is
spent under the barrier and the trajectory obeys Newton's law outside
the barrier) at points far behind the barrier we call Wigner time delay \cite{Wigner_1955}. 
In the following we will
apply
the Wigner time delay formalism to some exactly solvable basic systems under
the dynamics of the Schr{\"o}dinger equation with some potential $V(x)$.
These examples will give us some hints for the analysis of
tunnel-ionization process.

\subsection{Square potential}
\label{sec:Square_potential}

As a first example we consider the Wigner time delay during the
penetration of a wave packet through a box potential
\begin{equation}
  \label{eq:square_potential}
  V(x) = V_0 \left( \theta(x) - \theta(x-a) \right) 
\end{equation}
with $\theta(x)$ denoting the Heaviside step function.  The wave packet
propagating to the barrier and tunneling through it is constructed via
superposing the steady-state solutions with energy eigenvalues
$\varepsilon < V_0$,
\begin{equation}
  \label{square_potential_wave_packet_full}
  \braket{x|\psi(t)} = \int_0^{V_0} d \varepsilon \, \exp\left(- i
    \varepsilon (t-t_0) + i \phi(x,\varepsilon) \right) g(x,\varepsilon)
\end{equation}
with $g(x,\varepsilon) = \braket{\varepsilon|\psi(t_0)}
|u(x,\varepsilon)|$.  Here $\braket{\varepsilon|\psi (t_0)}$ is the
initial wave packet centered around $\varepsilon_0 < V_0$,
$|u(x,\varepsilon)|$ and $\phi(x,\varepsilon)$ are the amplitude and the
phase of the steady-state solution $u(x, \varepsilon)$ for $\varepsilon
< V_0$, respectively. Generally, the wave packet in
Eq.~(\ref{square_potential_wave_packet_full}) includes both transmitted
and reflected waves. To define the Wigner trajectory for the
transmitted wave packet, we omit the reflected wave packet from the
barrier and utilize the steady-state solutions with the positive current
\begin{equation}
  \label{positive_current_square_potential}
  u_{+}(x,\varepsilon) =
  \begin{cases}
    e^{i k_1 x} & x <0 \\
    C_1 \left(e^{-k_2 x}+i e^{k_2 x} \right) + C_2 \left(e^{-k_2 x}-i e^{k_2 x} \right) & 0\le x \le
a \\
    T e^{i k_1 x}  &  x > a \\
  \end{cases}
\end{equation} 
with $k_1 = \sqrt{2 \varepsilon}$, $k_2 = \sqrt{2 V_0 -2 \varepsilon}$,
and the matching coefficients $C_1$, $C_2$ and $T$
\footnote{\label{matching}Note that the matching coefficients are found
  requiring continuity of the wave function $u(x,\varepsilon)$ which
  includes the reflected portion of the wave function rather than $u_{+}
  (x, \varepsilon)$. Furthermore, one can also omit the reflected
  portion of the wave packet valid for the under-the-barrier motion by
  setting $C_2$ zero in
  Eq.~(\ref{positive_current_square_potential}). However it does not
  affect the Wigner time delay.}. The Wigner trajectory for the
transmitted wave can then be defined implicitly via
\begin{equation}
  \label{qm_tajectory_tunneling}
  \tau(x) =  \left. 
    \dfrac{\del \varphi_{+}(x,\varepsilon) }{\del \varepsilon}
  \right|_{\varepsilon=\varepsilon_0}
\end{equation}
where $\varphi_{+}(x,\varepsilon)$ is the phase of
$u_{+}(x,\varepsilon)$.

\begin{figure}
  \centering
  \includegraphics[scale=0.525]{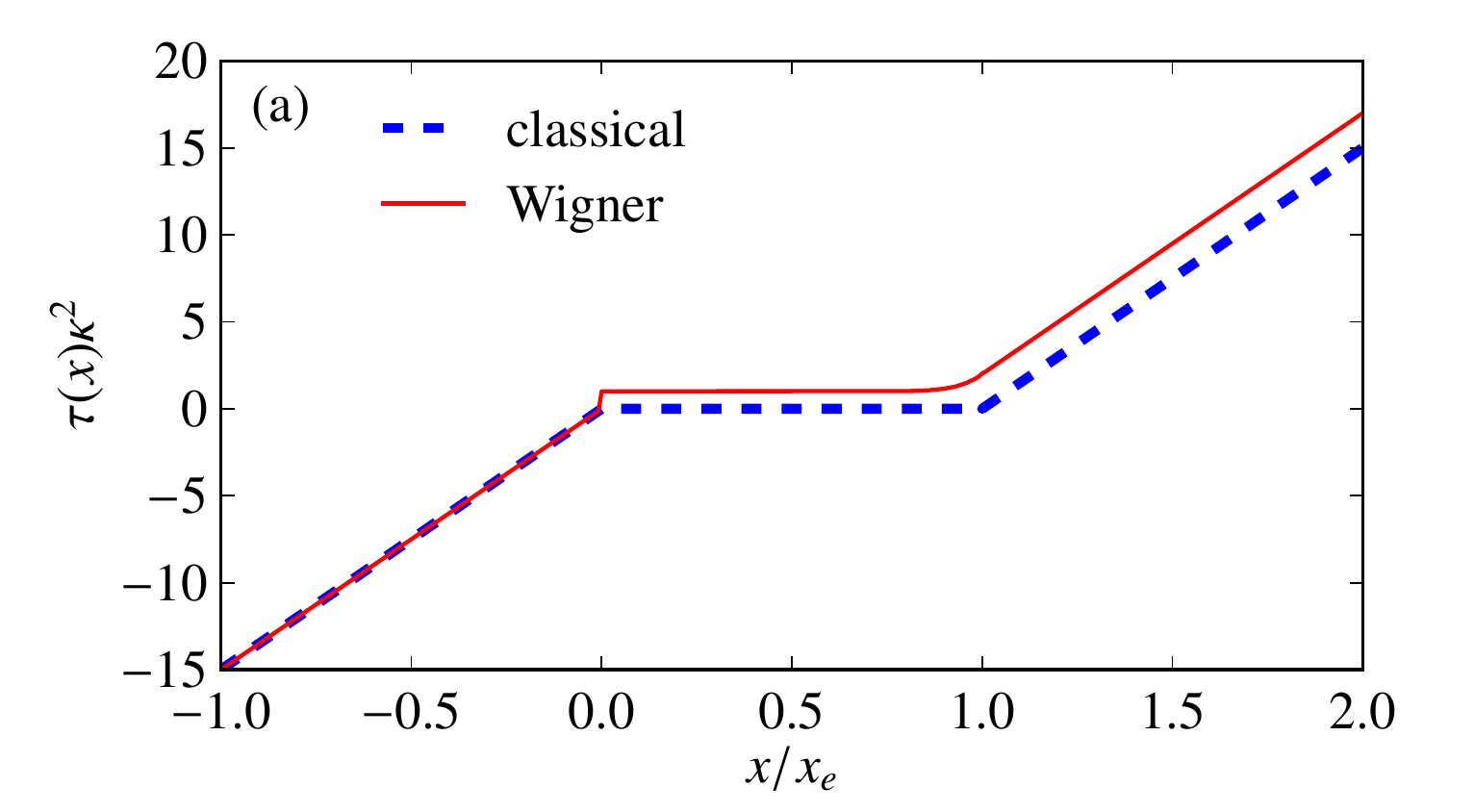}
  \includegraphics[scale=0.525]{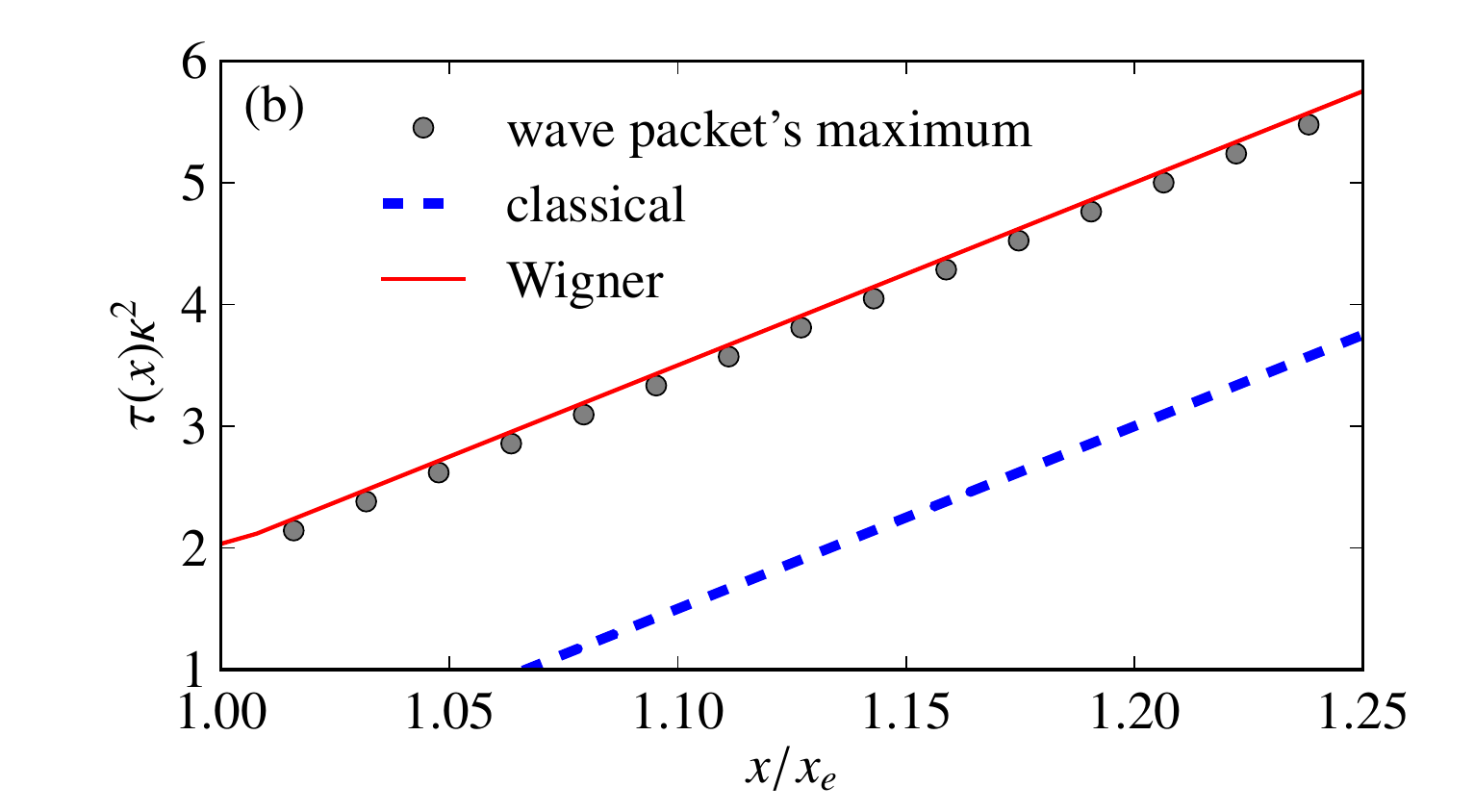}
  \caption{(Color online) (a) Comparison of the Wigner trajectory
    (\ref{qm_tajectory_tunneling}) (solid red line) and the classical
    (dashed blue line) trajectory for tunneling through a square
    potential barrier (\ref{eq:square_potential}).  The applied
    parameters are $V_0 = 2 \varepsilon_0$, $\varepsilon_0 = I_p$, $I_p
    = c^2-\sqrt{c^4-c^2 \kappa^2}$ with $\kappa=90$ and $a=14/\kappa$.
    (b) Close-up of sub-figure (a) with additionally showing the
    position of the wave packet's maximum.}
  \label{fig:square_potential_time_delay}
\end{figure}

The Wigner trajectory for the outgoing wave packet defined in
Eq.~(\ref{qm_tajectory_tunneling}), is shown
Fig.~\ref{fig:square_potential_time_delay}(a) and is compared with the
classical trajectory. From the latter one can see that the Wigner
trajectory, which initially coincides with the classical one, deviates
during tunneling from the classical trajectory. In the quasiclassical
limit $\kappa a \gg 1$, the Wigner time delay is
$\tau=1/2\sqrt{(V_0-\varepsilon_0)\varepsilon_0}$ which is formed at the
entering and exiting the barrier.

This simple example provides us the opportunity to compare Wigner's
approach with the exact trajectory of the wave packet's maximum which
can be calculated via Eq.~(\ref{square_potential_wave_packet_full}).
Let us choose the initial normalized wave packet in momentum space as
the Gaussian
\begin{equation}
  \braket{p|\psi(t_0 = 0)} = 
  \dfrac{e^{-i x_0 (p-p_0)}}
  {\left(2 \pi \delta p^2 \right)^{1/4}}
  \exp\left[-\left(\dfrac{p-p_0}{2 \delta p}\right)^2\right]\,,
\end{equation}
with the initial position $x_0$ and initial momentum $p_0 = \sqrt{2
  \varepsilon_0 }$. The position $x_0$ is assumed to be far away from
the barrier and the energy spread $\delta \varepsilon$ of the wave
packed so small that $\varepsilon_0 + \delta \varepsilon < V_0 $ with
$\delta \varepsilon=p_0\delta p$.  To be consistent with the
tunnel-ionization case where the wave packet has a sharp energy we
assume that $\delta p /p \ll 1$.  Then, the time evolution of the wave
packet becomes
\begin{equation}
  \braket{x|\psi(t)} = \int_0^{\infty} d \varepsilon \,
  \dfrac{\braket{x|\varepsilon}}{\left(2 \pi \delta p^2 \right)^{1/4}} e^{ - i
    \varepsilon t - i x_0 (\sqrt{2 \varepsilon}-\sqrt{2 \varepsilon_0})
    -\left(\frac{\sqrt{2 \varepsilon}-\sqrt{2 \varepsilon_0}}{2 \delta
        p}\right)^2
  }\,.
\end{equation} 
If we trace the maximum of the wave packet outside the barrier, we see
that it overlaps with the Wigner's trajectory defined in
Eq.~(\ref{qm_tajectory_tunneling}). The result can be seen in
Fig.~\ref{fig:square_potential_time_delay}(b).

To evaluate the role of the magnetic field in relativistic
tunnel-ionization, we modify the previous configuration by applying an
additional static magnetic field within the square potential. The vector
potential which generates the magnetic field $\vec{B}= -E_0 \left(
  \theta(x) - \theta(x-a) \right) \hat{y}$ may be written as
\begin{equation}
  \vec{A} = E_0 \left( x \left( \theta(x)-\theta(x-a)\right) + a
    \theta(x-a)\right) \hat{z}\,.
\end{equation}
Then, the Hamiltonian for this field configuration is
\begin{equation}
  H = \dfrac{p_x^2}{2} + \dfrac{[p_z + A_z (x)/c]^2}{2} + V(x)\,.
\end{equation}
Since the canonical momentum $p_z$ is conserved, $[p_z ,H] = 0$, the
energy eigenfunction has the form
\begin{equation}
  \braket{x,z|\varepsilon, p_z}= \dfrac{u(x,p_z,\varepsilon)}{\sqrt{2
      \pi}}e^{i  p_z z} ,
\end{equation} 
and the motion along $x$ coordinate is separable
\begin{equation}
  \label{sch_eq_for_sq_pot_with_B_field_in_x}
  \left[ 
    -\dfrac{1}{2} \dfrac{d^2}{d x^2} + 
    \dfrac{(p_z + A_z (x)/c)^2}{2} +
    V(x)
  \right] 
  u(x,p_z,\varepsilon) = \varepsilon \, u(x,p_z,\varepsilon)\,.
\end{equation}
The solution of Eq.~(\ref{sch_eq_for_sq_pot_with_B_field_in_x}) outside the
barrier is given by
\begin{equation}
  u(x,p_z,\varepsilon) =
  \begin{cases}
    u_1 (x,p_z,\varepsilon) = e^{i k_1 x}+R e^{-i k_1 x}  & \quad x <0 \\
    u_3 (x,p_z,\varepsilon) = T e^{i k_3 x}  & \quad x > a 
  \end{cases}
\end{equation}
with $k_1 = \sqrt{2 \varepsilon - p_z^2}$, $k_3 = \sqrt{2 \varepsilon -
  (p_z + a E_0 / c)^2}$, and reflection and transmission coefficients
are $R$ and $T$, respectively.  For the dynamics under the barrier ($0
\le x \le a$, $\varepsilon < V_0$) the Schr{\"o}dinger equation
\begin{equation}
  \left[ -\dfrac{1}{2}\dfrac{d^2}{d x^2} + \dfrac{(p_z - x E_0 / c )^2}{2} + V_0
  \right] u_2
  (x,p_z,\varepsilon)  = \varepsilon \, u_2 (x,p_z,\varepsilon)\,.
\end{equation}
has two linearly independent solutions which can be expressed using
parabolic cylinder function $D$ \cite{Abramowitz:Stegun:1972:Handbook}
as
\begin{multline}
  u_2 (x,p_z,\varepsilon) =
  C_1 \, D\left(
    -\frac{E_0+2 c V_0-2 c \varepsilon }{2 E_0}, 
    \frac{\sqrt{2} (c p_z +E_0 x)}{\sqrt{c E_0}}
  \right) \\
  + C_2 \, D\left(
    -\frac{E_0-2 c V_0+2 c \varepsilon }{2 E_0},
    \frac{i \sqrt{2} (c p_z +E_0 x)}{\sqrt{c E_0}}
  \right)\,.
\end{multline}
Here, the matching coefficients $C_1$ and $C_2$ can be found via using
the continuity of the wave function at the borders of the potential
$x=0$ and $x=a$.

The wave packet tunneling out of the potential barrier is 
\begin{multline}
  \braket{x,z|\psi(t)} = \int_{-\infty}^\infty d p_z \int_0^{V_0}
  d \varepsilon \,
  \braket{x,z|\varepsilon, p_z} \braket{\varepsilon, p_z|\psi(t)} \\
  = \dfrac{1}{\sqrt{2 \pi}} \int_{-\infty}^\infty d p_z \int_0^{V_0}
  d \varepsilon \, e^{ i z p_z - i \varepsilon (t-t_0) + i
    \phi_{+}(x,p_z,\varepsilon)
  } g(x,p_z, \varepsilon)\,,
\end{multline}
where $g(x,p_z, \varepsilon) = \braket{\varepsilon,p_z|\psi(t_0)}
|u(x,p_z,\varepsilon)|$, $\braket{\varepsilon,p_z|\psi (t_0)}$ is the
initial wave packet centered around ${p_z}_0$ and $\varepsilon_0$, and
$\phi_{+}(x,p_z,\varepsilon)$ is the phase of the outgoing part of
$u(x,p_z,\varepsilon)$.

Calculating the transition probability $|T|^2$ as a
function of $p_z$, we find that the maximum is reached at ${p_z}_0 =
-I_p / (2 c)$ which equals the kinetic momentum at the tunneling entry
$x=0$, $q_{z}(0)= -I_p / (2 c)$. At the tunneling exit $x=a$ the kinetic
momentum with maximal tunneling probability is $q_{z}(a) = {p_z}_0 +
A_z(a)/c=I_p / (2 c)$. 

\begin{figure}
  \centering
  \includegraphics[scale=0.7]{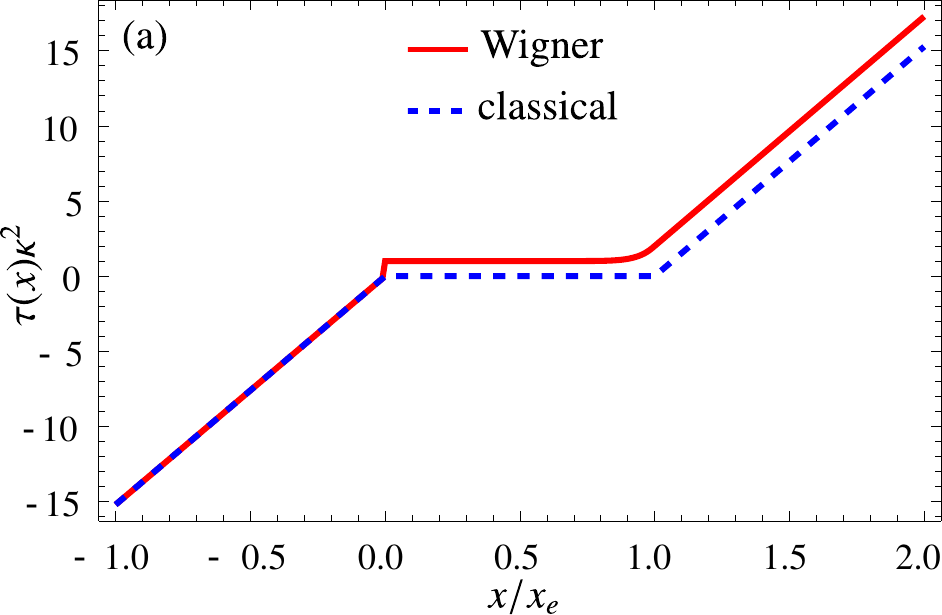}
  \\[3ex]
  \includegraphics[scale=0.7]{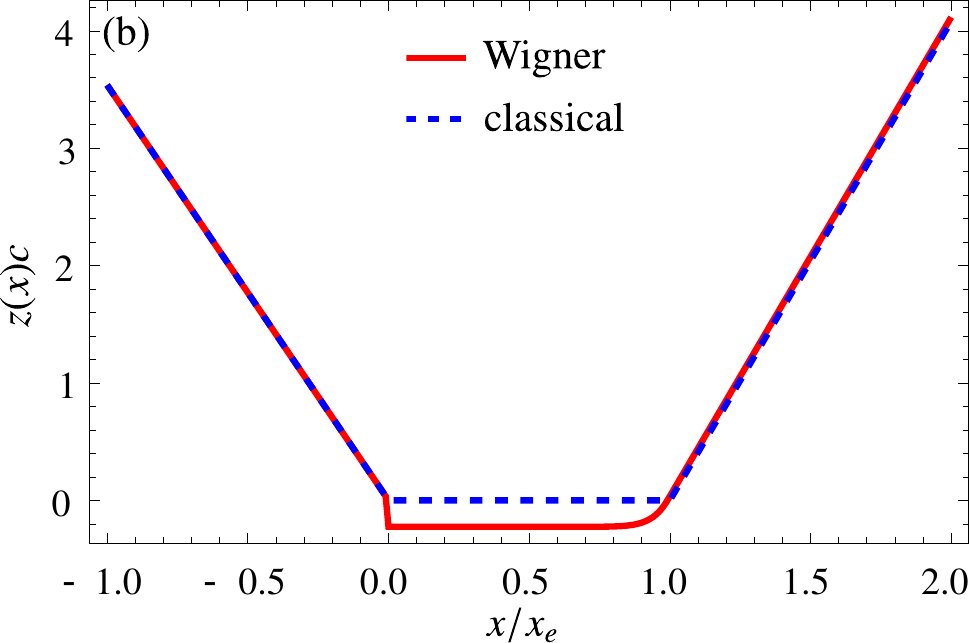}
  \caption{(Color online) Tunneling through a square potential with an
    additional magnetic field with $E_0 = \kappa^3/30$ and other
    parameters as in Fig.~\ref{fig:square_potential_time_delay}.
    Sub-figures (a)~and~(b) compare the Wigner (red solid lines)
    and the classical (blue dashed lines) trajectories along the
    $x$~direction and in the $x$-$z$-plane, respectively.}
  \label{square_potential_mag_time_delay}
\end{figure}

As the phase now depends also on $p_z$
Eq.~\eqref{qm_tajectory_tunneling} generalizes to
\begin{equation}
  \label{qm_tajectory_tunneling_with_mag}
  \tau(x) =  \left. 
    \dfrac{\del \varphi_{+}(x,p_z,\varepsilon) }{\del \varepsilon}
  \right|_{\varepsilon=\varepsilon_0}\,.
\end{equation}
Thus for each choice of $p_z$
Eq.~\eqref{qm_tajectory_tunneling_with_mag} defines a different
trajectory and setting $p_z={p_z}_0$ in
Eq.~(\ref{qm_tajectory_tunneling_with_mag}) gives the most probable
trajectory, which is shown in
Fig.~\ref{square_potential_mag_time_delay}(b) together with the
classical one.  Comparing Figs.~\ref{fig:square_potential_time_delay}(a)
and~\ref{square_potential_mag_time_delay}(a) shows that presence of the
magnetic field does not change the Wigner time delay.

In analogy to Eq.~(\ref{expectation_from_phase}), one can also define
the coordinate $z$ as a function of $x$
\begin{equation}
  \label{phase_drift}
  z = - \left. 
    \dfrac{\del \phi_{+} (x,p_z,\varepsilon_0) }{\del p_z}
  \right|_{p_z={p_z}_0}\,,
\end{equation}
which gives the most probable trajectory (path) in the $x$-$z$-plane as
shown in Fig.~\ref{square_potential_mag_time_delay}(b) together with the
corresponding classical trajectory.  Outside the barrier both are close
together; under the barrier, however, there is a clear spatial drift
into the $z$-direction, that is perpendicular to the tunneling
direction, of the Wigner trajectory as compared to the classical one.
Nevertheless, this spatial drift is (in contrast to the Wigner time
delay) not observable on the detector at remote distance.  This is
intuitively plausible, because the drift at the entry point to the
barrier $\Delta z_d(0)=q_{z}(0)\tau=-(I_p/2c)\tau$ is exactly
compensated by the drift at the exit point $\Delta
z_d(a)=q_{z}(a)\tau=(I_p/2c)\tau$.

\subsection{Linear potential}
\label{sec:Linear_potential}

\begin{figure}
  \centering
  \includegraphics[scale=0.7]{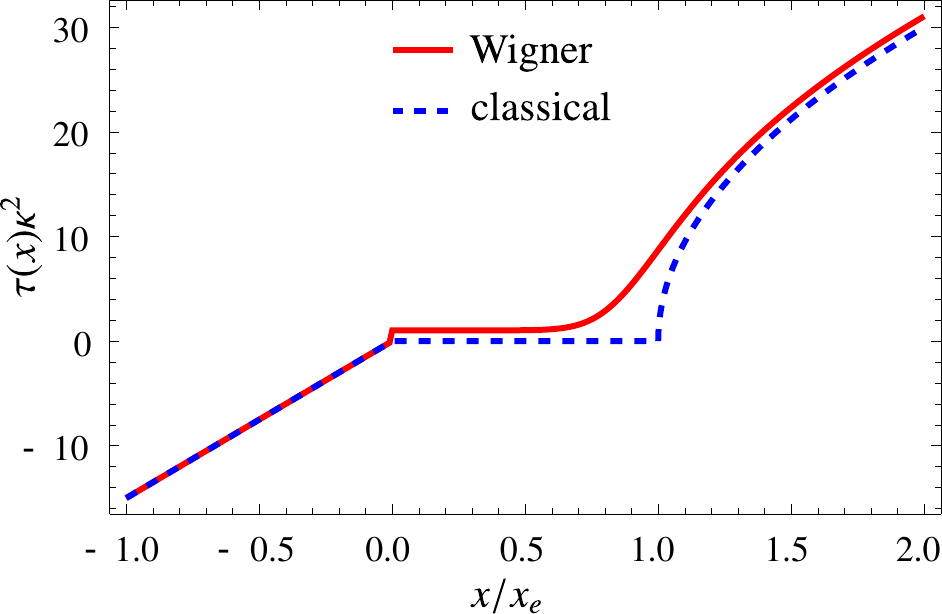}
  \caption{(Color online) Comparison of the Wigner
    trajectory (red solid line) and the classical trajectory (blue
    dashed line) for tunneling through a linear potential barrier
    \eqref{eq:linear_pot} for $V_0 = 2 \varepsilon_0 $,
    $\varepsilon_0 = I_p$, with the numerical parameters $\kappa = 90$, and $E_0/E_a=1/30$.}
  \label{lin_potential_time_delay}
\end{figure}

As a second example, we consider tunneling through a linear potential
barrier
\begin{equation}
  \label{eq:linear_pot}
  V(x) = \theta(x)\left(E_0 x + V_0 \right) \,.
\end{equation}
The solution of the
corresponding Schr{\"o}dinger equation is given for the domain $x<0$ by
\begin{subequations}
  \label{eq:linear_pot_psi}
  \begin{equation}
    u_1 (x,\varepsilon) = e^{i k_1 x}+R e^{-i k_1 x}
  \end{equation}
  with $k_1 = \sqrt{2 \varepsilon}$ and the reflection coefficient
  $R$. While in the region $x \ge 0 $, the solution yields
  \begin{multline}
    \label{airy_transmitted}
    u_2(x,\varepsilon) = \\
    T \left( \Ai\left(\frac{2 E_0 x+2 (V_0-\varepsilon )}{(2
          E_0)^{2/3}} \right) + i \Bi\left( \frac{2 E_0 x+2
          (V_0-\varepsilon )}{(2 E_0)^{2/3}}\right)
    \right) \\
    = T \Ai\left( e^{-2 \pi i/3}\frac{2 E_0 x+2 (V_0-\varepsilon )}{(2
        E_0)^{2/3}} \right)\,,
  \end{multline}
\end{subequations}
where $\Ai$ and $\Bi$ are the Airy function of the first and second
kind, respectively.  Under the barrier, that is $0\le x\le x_e$ with the
tunneling exit point $x_e = -\varepsilon_0 / E_0$, the wave function
(\ref{airy_transmitted}) is a supersposition of reflected and
transmitted portions.  The transmission coefficient $T$ in
\eqref{airy_transmitted} is deduced from matching the wave functions at
the border $x=0$. The phase of the total wave function
\eqref{eq:linear_pot_psi} is used to calculate the Wigner
trajectory \eqref{definition_of_time} which is shown in
Fig.~\ref{lin_potential_time_delay} in comparison with the classical
one.  This comparison shows that before but close to the tunneling exit
$x_e$ a substantial time delay builds up which is reduced after
tunneling.  Finally, a non-vanishing Wigner time delay remains which is
detectable at a remote detector. This time delay is induced at the
entering the barrier and equals by magnitude $\tau
=1/(2\sqrt{(V_0-\varepsilon_0)\varepsilon_0})$.

\subsection{Parabolic potential} 

\begin{figure}
  \centering
  \includegraphics[scale=0.7]{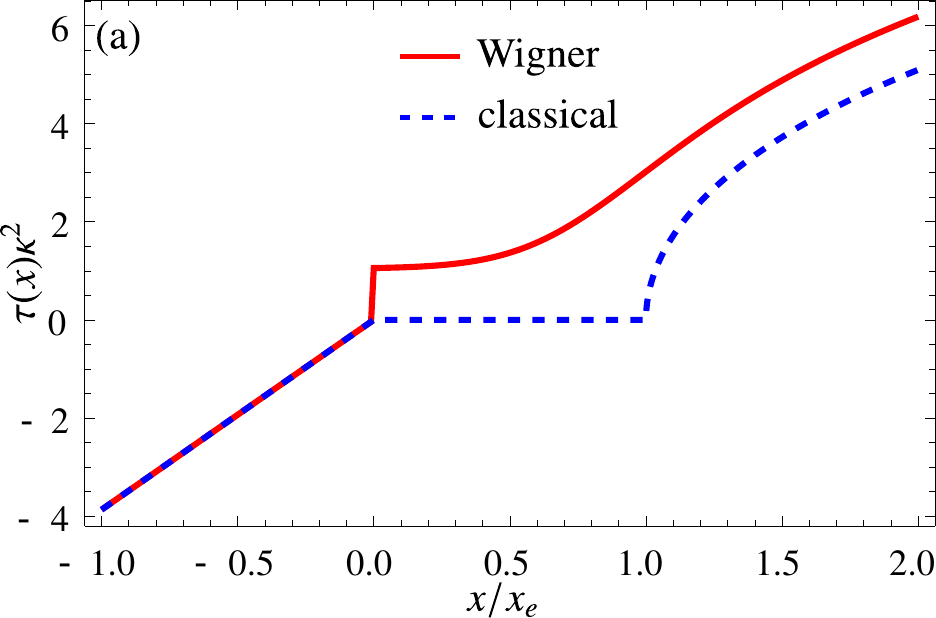}
  \\[3ex]
  \includegraphics[scale=0.7]{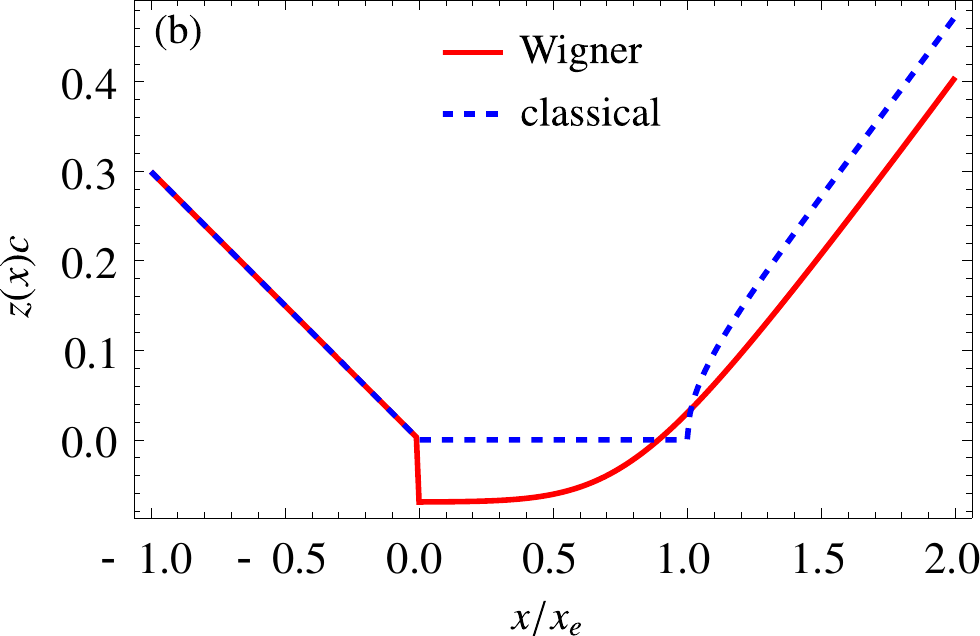}
  \caption{(Color online) (a) Comparison of the Wigner trajectory (red solid line) and the
classical
    trajectory (blue dashed line) for tunneling through the parabolic
    potential barrier \eqref{eq:parabolic_potential} for $V_0 = 2 \varepsilon_0 $, $\varepsilon_0
    = I_p$ with the numerical parameters
    $\kappa=90$, and $\beta = 1/30$. (b) The coordinate $z$
    as a function of $x$ for tunneling through a parabolic potential
    barrier in the presence of a magnetic field.}
  \label{parabolic_potential_time_delay}
\end{figure}

As a last example we examine the Wigner time delay for tunneling through
a parabolic potential barrier
\begin{equation}
  \label{eq:parabolic_potential} 
  V(x) = \theta(x)\left(-\beta \kappa^4 x^2 + V_0 \right)\,,
\end{equation}
with a dimensionless parameter $\beta$. In this case,
the exact solution for region $x \ge 0$ is given by 
\begin{equation}
  u_2 (x, \varepsilon) = 
  T \, D \left(
    -\frac{1}{2}-\frac{i (V_0-\varepsilon )}{\sqrt{2} \sqrt{\beta}
      \kappa ^2},-(-2)^{3/4} x \beta ^{1/4} \kappa
  \right)\,,
\end{equation}
with the transmission coefficient $T$ and $D$ denoting parabolic
cylinder functions \cite{Abramowitz:Stegun:1972:Handbook}.  The Wigner
and the classical trajectories for tunneling through the parabolic
potential are compared in Fig.~\ref{parabolic_potential_time_delay}(a).
Qualitatively the Wigner time delay behaves as for the linear potential.
In the barrier close to the exit a time delay is built up which is
reduced after tunneling and eventually a small non-vanishing Wigner time
delay remains.

In analogy to Sec.~\ref{sec:Square_potential} we add to the parabolic
potential \eqref{eq:parabolic_potential} a static magnetic field in the
region $x>0$ and investigate the spatial drift due to the magnetic field
in the Wigner time delay. Introducing the vector potential $A_z(x) =
\theta(x) E_0 x$ this scenario can be described by the Hamiltonian
\begin{equation}
  H = \dfrac{p_x^2}{2} + \dfrac{[p_z + A_z (x) /c]^2}{2} + V(x)\,.
\end{equation}
The canonical momentum that maximizes the tunneling probability equals
$p_z = - 0.15 I_p /c $. The coordinate $z$ as a function of $x$ is shown
in Fig.~\ref{parabolic_potential_time_delay}(b) for this canonical
momentum. As in the case of square potential with magnetic field, the
spatial shift between the classical and the Wigner trajectories is small.

\section{Time delay in tunnel-ionization} 
\label{sec:time_tunnel_ionization}

The previous sections' techniques can be also employed to analyze
tunnel-ionization in Hydrogenic ions as considered in
Sec.~\ref{sec:intuitive}.  The fundamental difference between the above
one-dimensional model systems and tunnel-ionization in Hydrogenic ions
is that in the former cases there is a source that produces a positive
current incident to the tunneling barrier.  In the latter case, however,
a bound state tunnels through a barrier and consequently the continuum
wave function has to be matched with the bound state wave function
(instead of the incident and the reflected plane waves as in the model
systems).  In the model systems of Sec.~\ref{sec:tunneling_time} the
Wigner trajectory and the classical trajectory coincide before they
enter the tunneling barrier. 

This, however, is no longer true for tunnel-ionization from bound
states, which can be explained by observing that some portion of the
bound state resides always in the barrier.  In order to define a
meaningful Wigner time delay one has to match the classical trajectory
with the Wigner one at the entry point $x_0$.  As a consequence, the
Wigner trajectory for tunnel-ionization $\tau_\mathrm{TI}(x)$ may be
defined via the relation
\begin{equation}
  \tau_\mathrm{TI}(x) =  \tau(x) - \tau(x_0)\,.
\end{equation}
Hence the corresponding the Wigner time delay for the tunnel-ionization
is identified as
\begin{equation}
  \tau_\mathrm{W} =  \tau_c (\infty) - \tau_\mathrm{TI}(\infty) \,
\end{equation}
with the classical trajectory $\tau_c (x)$. Since the coefficients that
match the bound wave function and the continuum wave function are
position independent the Wigner trajectory $\tau_\mathrm{TI}(x)$ for
tunnel-ionization is solely determined by the phase of the wave function
which is the solution of the corresponding Schr{\"o}dinger equation for the
potential
\begin{equation}
  \label{eq:tunneling_pot}
  V(x) = \theta(x-x_0) (x E_0 - \kappa/x ) 
\end{equation}
and represents asymptotically a plane wave.  Although, there is no
analytical solution for this potential approximate solutions can be
found in limiting cases which we will discuss in this section.

\subsection{Nonrelativistic case}
\label{sec:time_tunnel_ionization_nr}

\begin{figure}
  \centering
  \includegraphics[scale=0.7]{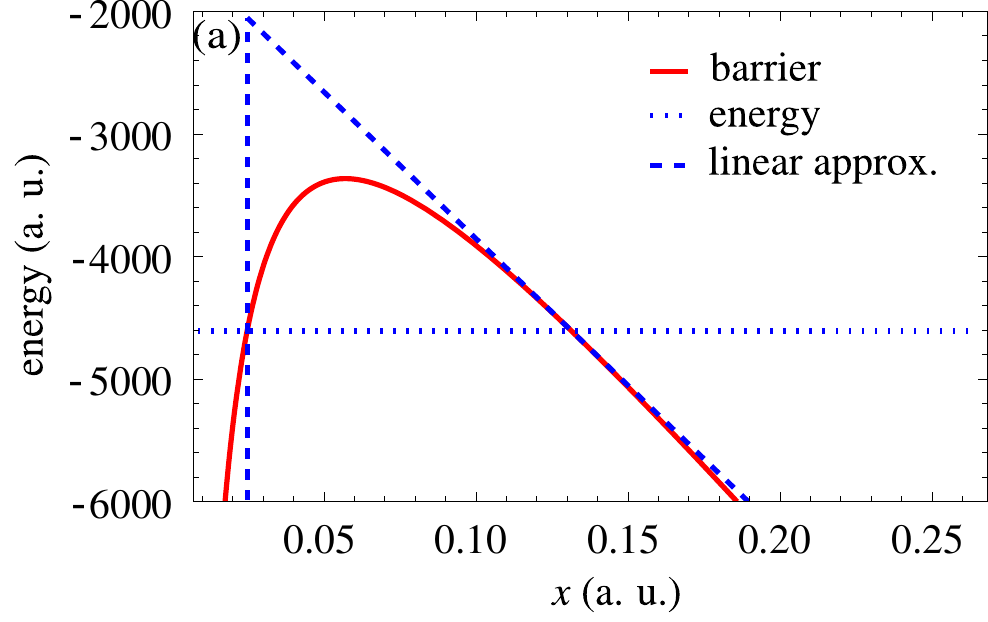}\\[3ex]
  \includegraphics[scale=0.7]{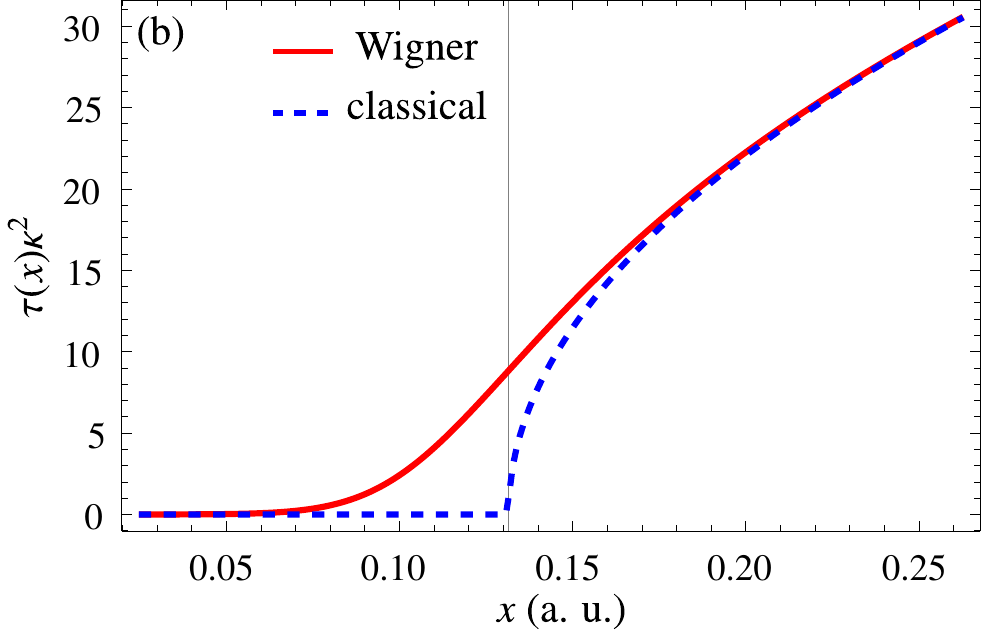}
  \caption{(Color online) (a) The potential barrier
    \eqref{eq:tunneling_pot} (red solid line) and its linear
    approximation \eqref{eq:linear_approx} (blue dashed line) in the
    deep-tunneling regime. The blue dotted line indicates the energy
    level of the bound state with $\varepsilon = -I_p$.  (b) The Wigner
    trajectory (red solid line) and the classical trajectory
    (blue dashed line) in the deep-tunneling regime for nonrelativistic
    tunnel-ionization with $\kappa = 90$ and $E_0/E_a = 1/30$. Vertical
    black line indicates the exit coordinate.}
  \label{non_rel_ti_dt_td}
\end{figure}
In the one-dimensional tunneling picture, the relevant Schr{\"o}dinger
equation with the electric dipole approximation is given by
\begin{equation}
  \left(
    -\dfrac{1}{2} \dfrac{d^2}{d x^2} + x E_0 - \dfrac{\kappa}{|x|} 
  \right) \psi(x) = \varepsilon \, \psi(x)\,.
\end{equation} 
When the tunneling-potential is of sufficient height (deep-tunneling
regime) the potential barrier \eqref{eq:tunneling_pot} can be
approximated near the tunneling exit point $x_e$ by a linear potential
\begin{equation}
  \label{eq:linear_approx}
  V(x) = V(x_e) + V'(x_e) (x-x_e)\,,
\end{equation}
see Fig.~\ref{non_rel_ti_dt_td}(a).  With this approximation the problem
of tunnel-ionization in Hydrogenic ions resembles the case of a linear
potential as discussed in Sec.~\ref{sec:Linear_potential}.  Therefore,
in the deep-tunneling regime, the solution of the Schr{\"o}dinger equation
is given by the Airy function
\begin{equation}
  \psi(x) = 
  \Ai\left(
    e^{-2 \pi i/3} \frac{-2 x V'(x_e)-2 \left(
        \varepsilon+ V(x_e)-x_e V'(x_e)
      \right)}{2^{2/3}
      \left(-V'(x_e) \right)^{2/3}}
  \right)\,.
\end{equation}
Note, that the potentials considered in Sec.~\ref{sec:Linear_potential}
and here differ in their positions, heights, and slopes.  The comparison
between the Wigner trajectory and the classical trajectory are plotted
in Fig.~\ref{non_rel_ti_dt_td}(b). In contrast to the linear potential
case of Sec.~\ref{sec:Linear_potential} the Wigner trajectory catches up
the classical one at far distance and consequently the Wigner time delay
vanishes.

\begin{figure}
  \centering
  \includegraphics[scale=0.7]{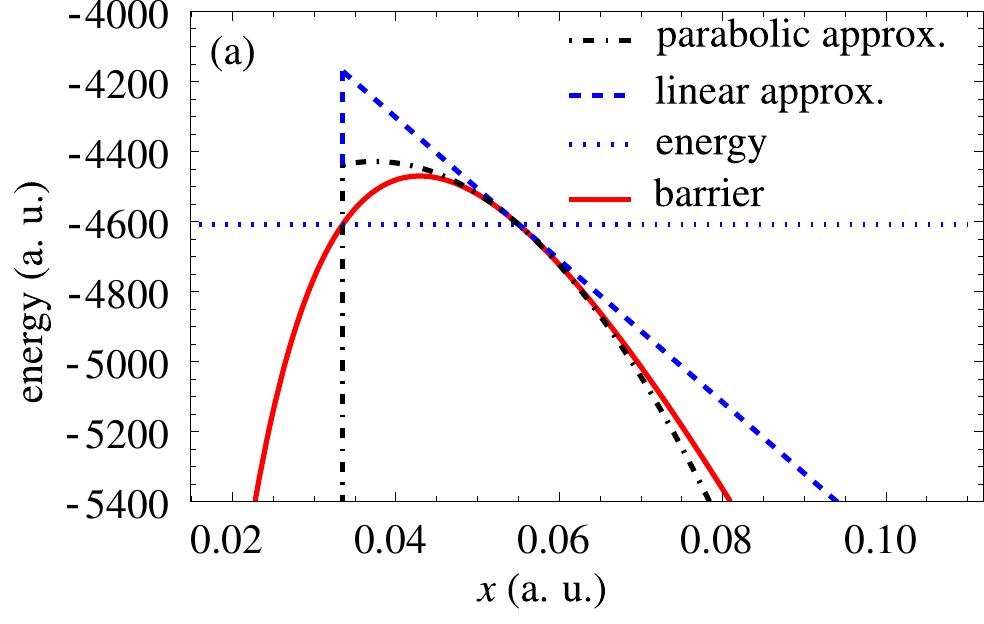}
  \\[3ex]
  \includegraphics[scale=0.7]{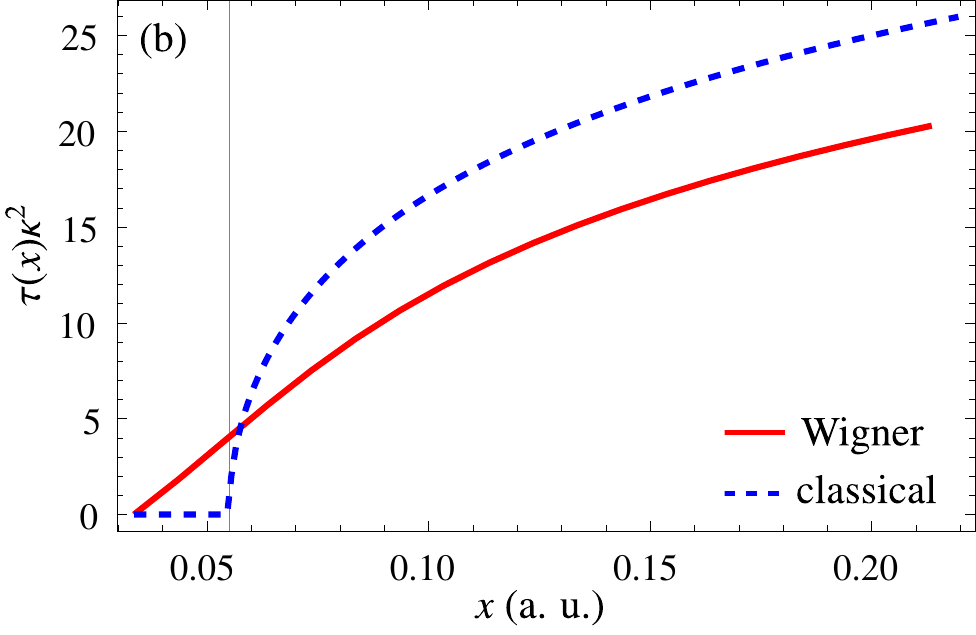}
  \caption{(Color online) (a) The tunneling barrier
    \eqref{eq:tunneling_pot} (red solid line), its linear approximation
    \eqref{eq:linear_approx} (blue dashed line), and the quadratic
    approximation \eqref{eq:quadratic_approx} (black dash-dotted line)
    in the near-threshold-tunneling regime. The blue dotted line
    indicates the energy level of the bound state with $\varepsilon =
    -I_p $.  (b) The Wigner (red solid line) and the classical
    trajectories (blue dashed line) in the near-threshold-tunneling
    regime for nonrelativistic tunnel-ionization with $\kappa =90$ and
    $E_0/E_a=1/17$. Vertical black line indicates the exit coordinate.}
  \label{non_rel_ti_ch_td}
\end{figure}

When the electric field strength is increased but the dynamics still
remains in the tunneling regime the linear approximation
\eqref{eq:linear_approx} becomes invalid. We may call this regime
near-threshold-tunneling regime of ionization. In this regime the
potential may be approximated by including the next quadratic term
\begin{equation}
  \label{eq:quadratic_approx}
  V(x) = V(x_e) + V'(x_e) (x-x_e) + V''(x_e)\dfrac{(x-x_e)^2}{2}\,,
\end{equation}
see Fig.~\ref{non_rel_ti_ch_td}(a).  As a consequence, the solution of
the Schr{\"o}dinger equation in the near-threshold-tunneling regime is given
in the form of parabolic cylinder function as
\begin{subequations}
  \begin{align}
    \label{D_function_for_non_rel}
    \psi(x) & =  D \left(a,b\right)\\
    \intertext{with}
    a & = -\frac{i\left(V'(x_e)^2-\left(2 \varepsilon +2
          V(x_e)+i\sqrt{V''(x_e)}\right)
        V''(x_e)\right)}{2V''(x_e)^{3/2}}\,,\\
    b &= \frac{(1-i)
      \left(V'(x_e)+(x-x_e)V''(x_e)\right)}{V''(x_e)^{3/4}}\,.
  \end{align}
\end{subequations}
Comparison of the Wigner and the classical trajectories in the
near-threshold-tunneling regime is shown in
Fig.~\ref{non_rel_ti_ch_td}(b). A non-vanishing Wigner time delay exists
in this case due to the parabolic character of the potential barrier
near the tunneling exit $x_e$.


\subsection{Magnetic dipole effects}

When the laser's magnetic field is taken into account, tunnel-ionization
in Hydrogenic systems can be described by a one-dimensional model if one
introduces a position dependent energy level inside the barrier as
discussed in Sec.~\ref{sec:intuitive}. In this one-dimensional model the
role of the curvature of the potential barrier and, therefore, its
approximations are the same as in the nonrelativistic case within the
electric dipole approximation. As for the square and the parabolic
potentials with magnetic field we calculate the Wigner time delay when
the tunneling probability is maximal. This happens at a certain
non-vanishing momentum along the laser's propagation direction.

The leading term in $1/c$ is the magnetic dipole correction. Including
the latter into the Schr{\"o}dinger equation in the electric dipole
approximation yields\pagebreak[1]
\begin{equation}
  \left[
    -\dfrac{1}{2}\dfrac{d^2}{d x^2} + 
    \dfrac{(p_z - x E_0 /c)^2}{2}
    + x E_0 - \dfrac{\kappa}{|x|}
  \right] \psi (\vec{x}) = \varepsilon \psi(\vec{x})\,.
\end{equation}
As a consequence of the presence of the vector potential term, for both
deep-tunneling and near-threshold-tunneling regimes, the relevant
solutions are given by parabolic cylinder functions. Since the quadratic
approximation covers also the deep-tunneling regime, in the former case
the solution yields
\begin{subequations}
  \begin{equation}
    \label{D_function_for_non_rel_mag}
    \psi(x) =  D \left(a,b\right)
  \end{equation}
  with
  \begin{multline}
    a = \frac{c \left(-c V''(x_e) \sqrt{c^2 V''(x_e)+E_0^2}-E_0^2
        \sqrt{{E_0^2}/{c^2}+V''(x_e)} \right)}{2 \left(c^2
        V''(x_e)+E_0^2\right)^{3/2}} \\
    + \frac{c^3 \left(V''(x_e) \left(p_z^2+2 V(x_e)-2 \varepsilon
        \right)-V'(x_e)^2\right) }{2 \left(c^2
        V''(x_e)+E_0^2\right)^{3/2}} \\
    + \frac{c \left(-2 c E_0 p_z \left(V'(x_e)-x_e V''(x_e)\right)
      \right)}{2 \left(c^2
        V''(x_e)+E_0^2\right)^{3/2}} \\
    + \frac{c \left( E_0^2 \left(x_e^2 V''(x_e)-2 \left(x_e
            V'(x_e)+\varepsilon \right)+2 V(x_e)\right)\right)}{2
      \left(c^2 V''(x_e)+E_0^2\right)^{3/2}} \,,
  \end{multline}
  \begin{equation}
    b =  -\frac{i \sqrt{2} \left(c^2 (x-x_e) V''(x_e)+c^2 V'(x_e)+E_0 (c
        p_z+E_0 x)\right)}{c^2 \left({E_0^2}/{c^2}+V''(x_e)\right)^{3/4}} \,.
  \end{equation}
\end{subequations}

\begin{figure}
  \centering
  \includegraphics[scale=0.7]{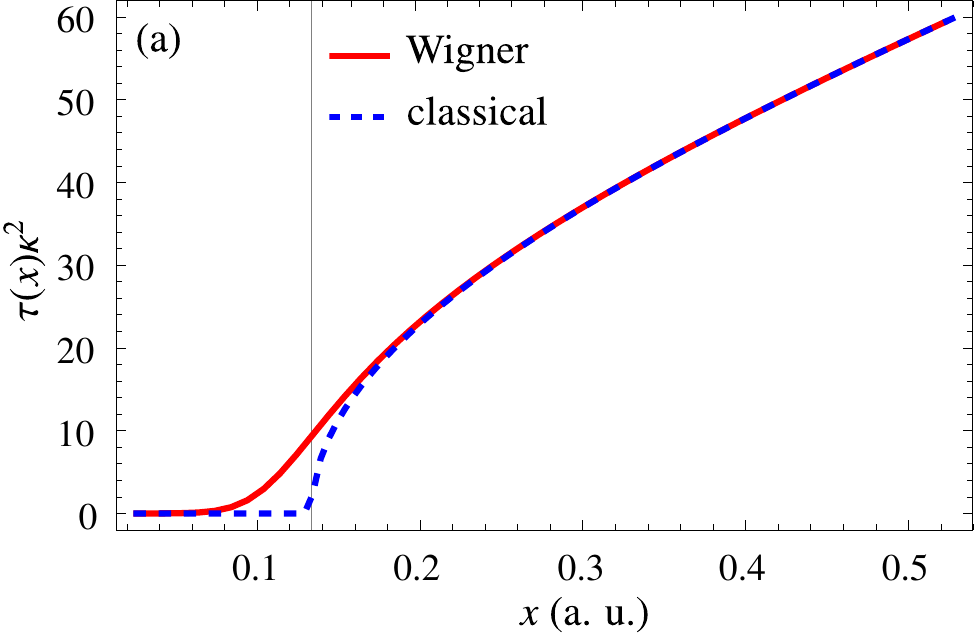}\\[3ex]
  \includegraphics[scale=0.7]{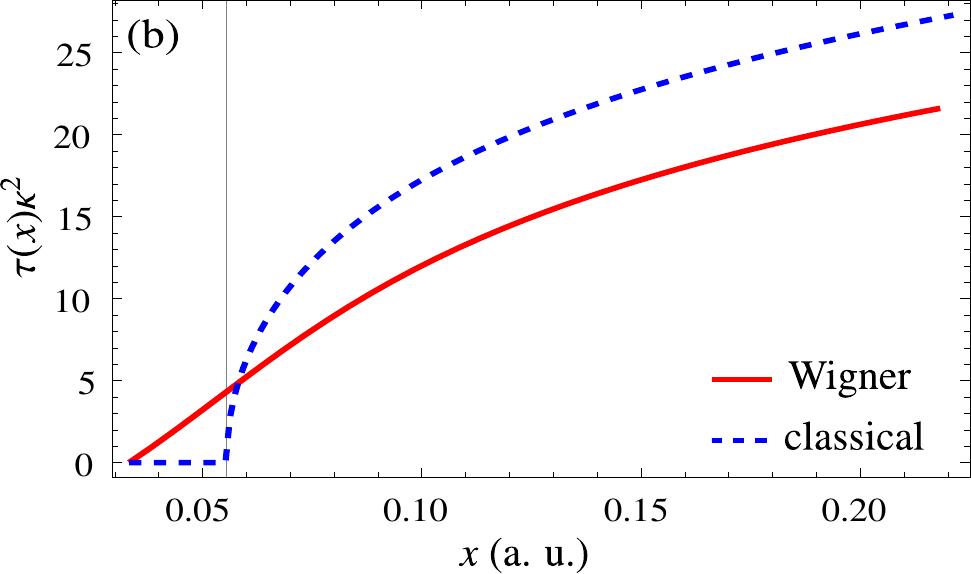}\\[3ex]
  \includegraphics[scale=0.7]{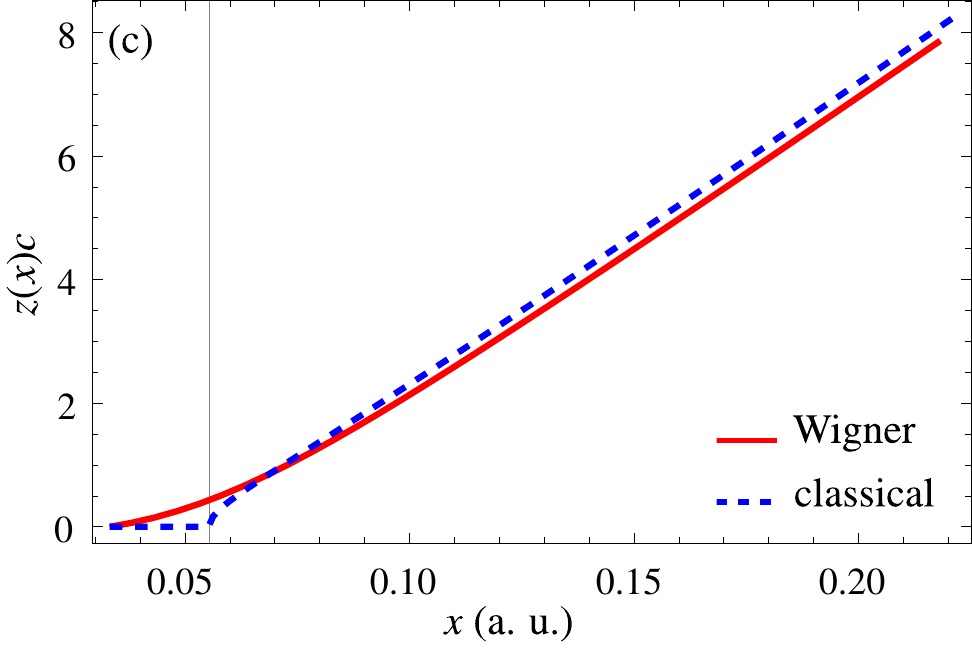}
  \caption{(Color online) Comparison of the Wigner (red solid line) and the
    classical trajectories (blue dashed line) for tunnel-ionization
    taking into account leading effects in $1/c$ (magnetic dipole
    effects): (a) for the deep-tunneling regime with the most probable
    momentum at the exit $q_z (x_e) = 0.28 I_p /c $, $\kappa = 90$ and
    $E_0/E_a=1/30$; (b) for the near-threshold-tunneling regime with $q_z (x_e)
    = 0.12 I_p /c $, $\kappa = 90$ and $E_0/E_a=1/17$.  Corresponding
    classical and Wigner trajectories $z$ as a function of $x$ are
    presented in sub-figure (c) for parameters of sub-figure
    (b). Vertical black lines indicate the exit coordinate.}
  \label{rel_ti_dt_td}
\end{figure} 

In the deep-tunneling regime, where the potential barrier is
approximately linear, the Wigner time delay vanishes as plotted in
Fig.~\ref{rel_ti_dt_td}(a). However, when tunneling happens in the
near-threshold-tunneling regime, there exists a non-zero Wigner time
delay, see Fig.~\ref{rel_ti_dt_td}(b).  Its order of magnitude equals
the result of the nonrelativistic near-threshold-tunneling regime
ionization, compare Figs.~\ref{non_rel_ti_ch_td}(b)
and~\ref{rel_ti_dt_td}(b).  Moreover, due to the non-vanishing Wigner
time delay, a spatial shift between the classical trajectory and the
Wigner trajectory along the laser's propagation direction is expected
because of the Lorentz force.  Both trajectories are shown in
Fig.~\ref{rel_ti_dt_td}(c).

\subsection{Relativistic effects}

\begin{figure}
  \centering
  \includegraphics[scale=0.7]{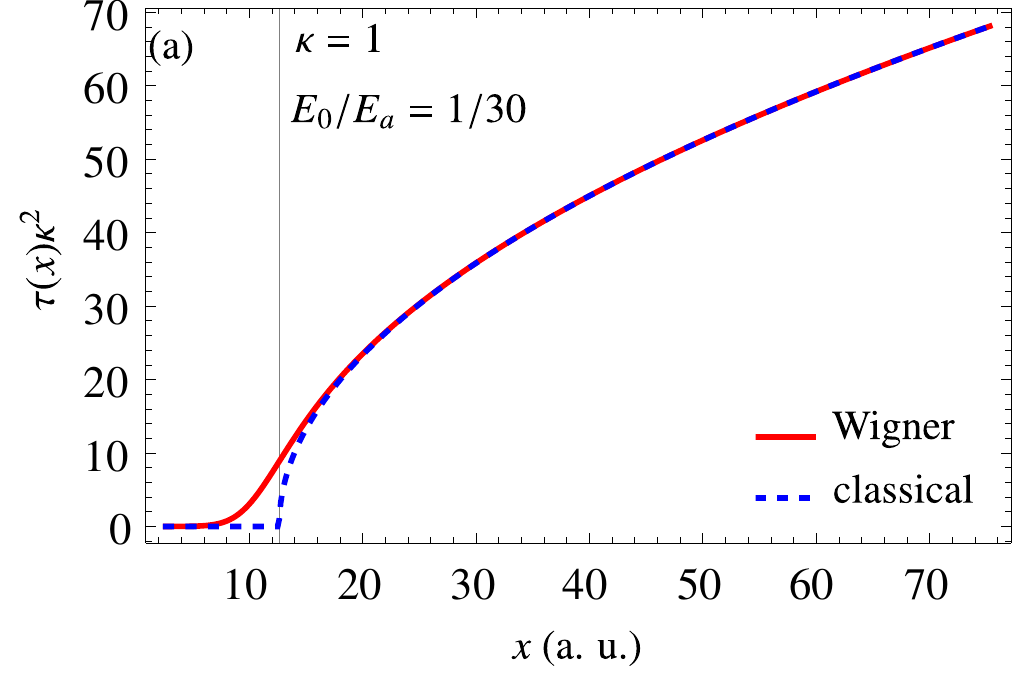}
  \\[3ex]
  \includegraphics[scale=0.7]{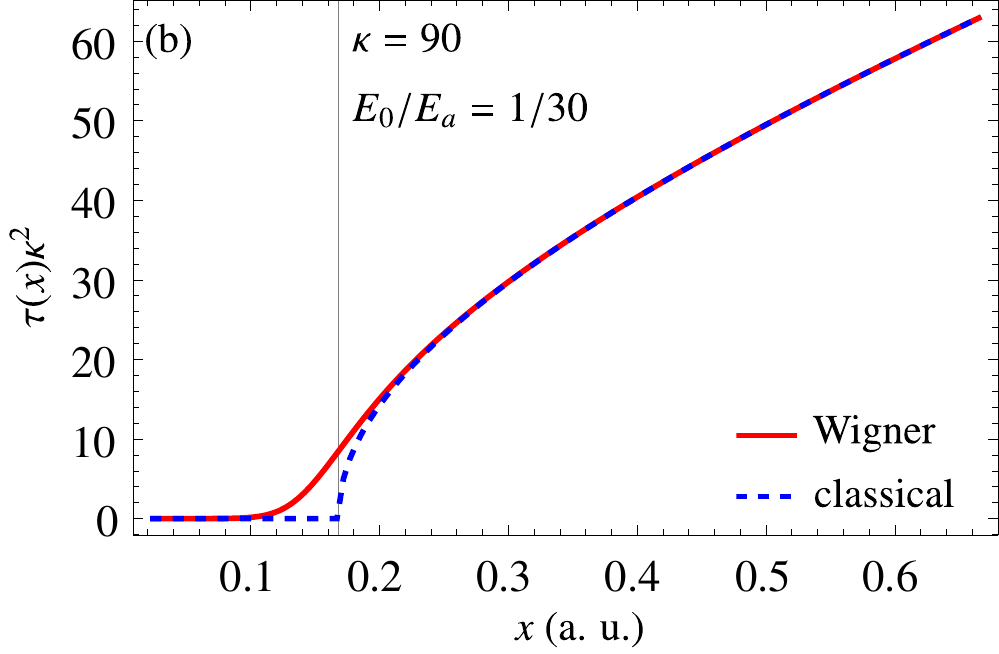}
  \caption{(Color online) Comparison of the Wigner trajectory (red solid
    line) and the classical trajectory (blue dashed line) for tunnel
    ionization taking into account kinematic relativistic effects for
    the deep-tunneling regime employing nonrelativistic (part (a)) and
    relativistic (part (b)) parameters.  In both parameter regimes no
    non-zero time delay is detectable at remote distance.  Vertical
    black line indicates the exit coordinate.}
  \label{rel_lin_deep_k30}
\end{figure}

In order to investigate relativistic effects in the relevant weakly
relativistic regime, the leading relativistic corrections to the kinetic
energy are expected to be dominant with regard to the Wigner time
delay. For simplicity, the exact Klein-Gordon equation is solved where,
however, the higher order relativistic effects play no significant role
in our scenario. Other leading relativistic effects in $1/c^2$ as those
dependent on the spin and on the magnetic field are conjectured to be
smaller than the leading relativistic correction to the kinetic
energy. In our one-dimensional intuitive picture of
Sec.~\ref{sec:intuitive}, the relevant corresponding Klein-Gordon
equation yields
\begin{equation}
  \left( - c^2 \dfrac{d^2}{d x^2} + c^4 \right) \psi(x) =
  \left(\varepsilon - V(x)\right)^2  \psi(x) \,.
\end{equation}
with the potential barrier (\ref{eq:tunneling_pot}). 

In the deep-tunneling regime, we can linearize the potential barrier,
i.\,e., $V(x)=V(x_e) + V'(x_e)(x-x_e)$.  The corresponding solution
which has asymptotically the form of a plane wave reads
\begin{subequations}
  \label{D_function_for_rel}
  \begin{equation}
    \psi(x) = D \left(a,b\right)
  \end{equation}
  with\pagebreak[1]
  \begin{align}
    a & = -\frac{1}{2}+\frac{i c^3}{2 V'(x_e)} \, ,\\
    b &=  -\frac{(1+i) \left((x-x_e) V'(x_e)+V(x_e)-\varepsilon \right)}{\sqrt{c}
      \sqrt{V'(x_e)}} \,.
  \end{align}
\end{subequations}
For the deep-tunneling regime, where the Wigner time delay is zero in
the nonrelativistic case, the relativistic corrections to the kinetic
energy are also not able to induce a non-zero Wigner time delay as
illustrated in Fig.~\ref{rel_lin_deep_k30}.

\begin{figure}
  \centering
  \includegraphics[scale=0.68]{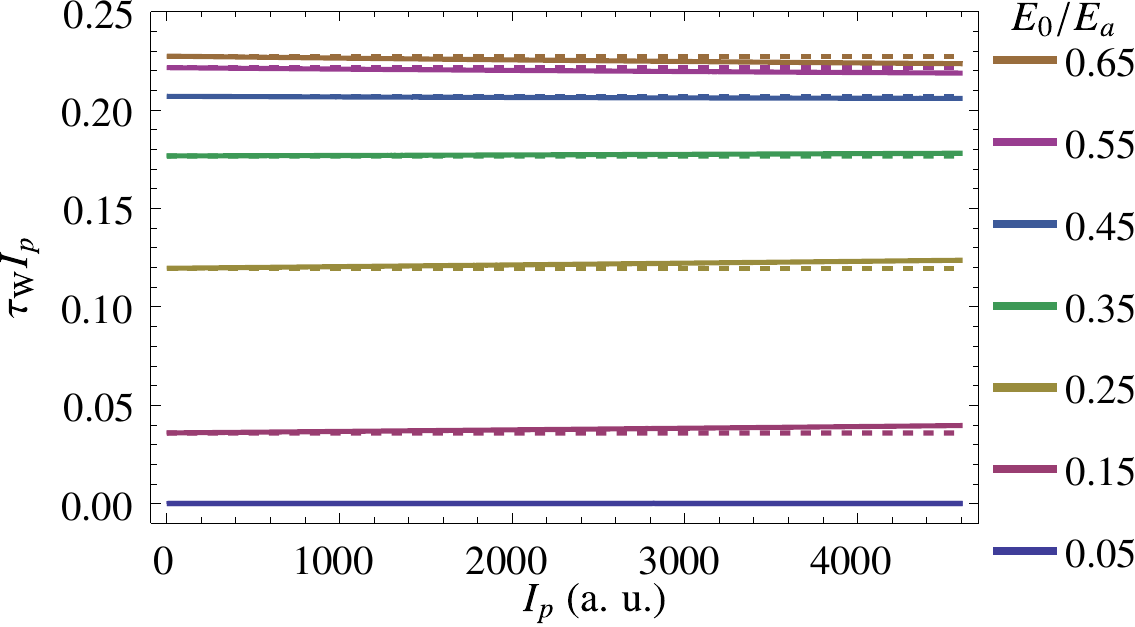}%
  \caption{(Color online) The scaled Wigner time delay $\tau_\mathrm{W}
    I_p$ as a function of $I_p$ for different values of $E_0/E_a$ for
    tunnel-ionization taking into account relativistic kinematic effects
    (solid lines) and for nonrelativistic tunnel-ionization (dashed
    lines) for the zero-range potential. Note the increasing small deviations of the 
scaled Wigner time delay with increasing $I_p$(relativistic effects).}
  \label{rel_time_delay_Ip}
\end{figure}

As pointed out in Sec.~\ref{sec:time_tunnel_ionization_nr} an analysis
of tunneling from a Coulomb potential in the near-threshold-tunneling
regime employs best a quadratic fitting of the tunneling barrier in the
vicinity of the tunneling exit. However, there is still no analytic
solution to this problem. Therefore, we replace now the Coulomb
potential again with a zero-range potential rendering the linear
approximation applicable again apart for the singular position of the core.
Note that the high nonlinearity of the effective potential barrier near the core is still
maintained. Then, the solution indicates that the
leading relativistic correction to the kinetic energy has a negligible
effect on the Wigner time delay as illustrated in
Fig.~\ref{rel_time_delay_Ip}. In this figure, the scaled Wigner time delay
$\tau_\mathrm{W} I_p$ for the nonrelativistic (dashed lines) as well
as relativistic case (solid lines) is shown for different values of
$I_p$ and of $E_0/E_a$.

\subsection{Intuitive explanation of the Wigner time delay}
\label{ssc:intuitive_time_delay}

Finally, we address the formal definition of deep-tunneling and
  near-threshold-tunneling and clarify the reason of the differing
  behavior of the Wigner time delay in these regimes.  A general rule
  for the validity of the linear approximation, which allows us to
  neglect the quadratic and the higher-order terms in the expansion of
  the potential barrier, can be given via the condition
  \begin{equation}
    \label{deep_tunneling_condition}
    \left| \frac{V''(x_e)}{V'(x_e)} \, \delta x \right| \ll 1 
  \end{equation}
  on the characteristic distance $\delta x$ that will be quantified later.
  Indeed, this condition defines the deep-tunneling regime. The regime, which
  meets the condition
  \begin{equation}
    \label{deep_tunneling_condition_1}
    \left| \frac{V''(x_e)}{V'(x_e)} \, \delta x \right| \sim 1 
  \end{equation}
  is identified as the near-threshold-tunneling regime.  In the latter
  case the potential has to be approximated by including also at least
  the quadratic term
  which is sufficient unless it is too close to the threshold
  \begin{equation}
    V(x) = V(x_e) + V'(x_e) (x-x_e) + V''(x_e)\dfrac{(x-x_e)^2}{2}\, .
  \end{equation}

  In Sec.\ VII we characterized the Wigner time delay as the asymptotic
  deviation of the maximum of the wavepacket position from its
  corresponding classical path. Therefore, the typical distance $\delta
  x$ can be identified with the deviation of the maximum of the
  wavepacket position with respect to the classical one, viz.\ $x = x_e
  + \delta x$, and similarly for the momentum $p = p_e + \delta p$, with
  $x_e$ and $p_e$ denoting the classical position and momentum at the
  tunnel exit.  The typical distance $\delta x$ may be estimated by
  considering the Hamiltonian for a nonrelativistic particle which
  tunnels trough a potential barrier $ V(x)$.  The corresponding
  Schr{\"o}dinger Hamiltonian
  \begin{equation}
    H = \frac{p^2}{2} + V(x) 
  \end{equation}
  can be written in terms of $\delta x$ and $\delta p$ as
  \begin{equation}
    \frac{\delta p^2}{2} + \delta x \, V'(x_e) = 0 \, ,
  \end{equation}
  with $V(x_e) = -I_p$ and $p_e = 0$.  Employing the uncertainty
  relation $\delta x \, \delta p \sim 1$, the deviation from the
  classical trajectory is obtained as
  \begin{equation}
    \label{estimation_of_delta_x}
    \left| \delta x \right|  \sim \left| V'(x_e)^{-1/3} \right| \,.
  \end{equation}
  Inserting Eq.~(\ref{estimation_of_delta_x}) into
  Eq.~(\ref{deep_tunneling_condition}), the condition for the validity of
  the linear approximation reads
  \begin{equation}
    \label{deep_tunneling_condition_2}
    \left| \frac{V''(x_e)}{V'(x_e)^{4/3}} \right| \ll 1 .
  \end{equation}
  The latter condition which also quantifies the transition regime
  between the deep-tunneling and the near-threshold-tunneling
  regime can be also expressed via the parameter $E_0/E_a$ in the case
  of the one-dimensional potential~(111) as
  \begin{equation}
    \label{deep_tunneling_condition_3}
    \left(\frac{16 E_0}{E_a}\right)^{5/3} \ll 1 \, .
  \end{equation}
  For instance, $(16 E_0/E_a)^{5/3} \approx 0.3$ for the deep tunneling
  regime with $E_0/E_a = 1/ 30$, whereas $(16 E_0/E_a)^{5/3} \approx 0.9$ for
  the near-threshold-tunneling regime with $E_0/E_a = 1/ 17$. In the case of
  a short-range atomic potential, where $V''(x_e)=0$, the condition of
  the near-threshold-tunneling regime,
  Eq. (\ref{deep_tunneling_condition_2}), should be modified. The
  tunneling potential is not linear in this case due to the edge of the
  triangular shaped effective barrier which becomes essential for the
  tunneling time at $\delta x \sim x_e$.  Therefore, in the case of a
  short-range atomic potential Eq. (\ref{deep_tunneling_condition_3}) is
  replaced by $\delta x/x_e\sim (E_0/E_a)^{2/3}\ll 1$.

 It should be noted here that the nonrelativistic Schr{\"o}dinger equation
  \begin{equation}
    \left(
      -\dfrac{1}{2} \vec{\nabla}^2 + x E_0 - \dfrac{\kappa}{r} 
    \right) \psi(\vec{x}) = \varepsilon \, \psi(\vec{x})
  \end{equation}
  can be separated in cylindrical parabolic coordinates such that the
  three-dimensional problem reduces effectively to a one-dimensional one
  with the one-dimensional potential for the ground state\cite{Landau_3,Pfeiffer_2012a}
  \begin{equation}
    \label{eq:parabolic_pot}
    V(\zeta) = 
    - \frac{1}{4 \zeta} - \frac{1}{8 \zeta^2} - \frac{1}{8} E_0 \zeta \, .
  \end{equation}
  A calculation of the tunneling time delay utilizing the potential
  (\ref{eq:parabolic_pot}) instead of (\ref{eq:tunneling_pot})
  would give qualitatively the same results in both tunneling regimes
  because both potentials have the same behavior in the continuum range
  of the potential barrier. The quantitative difference comes only from
  a numerical value in the transition regime between the deep-tunneling
  and the near-threshold-tunneling regime. Namely, the
  condition~(\ref{deep_tunneling_condition_3}) can be written for the
  potential~(\ref{eq:parabolic_pot}) as
  \begin{equation}
    \left(\frac{9 E_0}{E_a}\right)^{5/3} \ll 1 \, 
  \end{equation}
  where $E_0/E_a = 1/9$ is in the border between tunnel-ionization
  and over-the-barrier ionization.

Further, this classification of tunneling regimes allows to formulate a condition for a non-zero
Wigner time delay.  When the potential barrier is linear on the typical distance $\delta x$ around
the exit, i.e., in the deep tunneling regime, the time delay vanishes at far distances. If this is
not the case, in the near-threshold-tunneling regime, a non-zero tunneling time is expected. 
This is consistent with our results in Fig.~\ref{rel_time_delay_Ip}.

To sum up, a possible experimental verification of the tunneling time
delay is expected to be feasible in the near-threshold-tunneling
regime. This delay is approximately proportional to $1/I_p$ as shown in
Fig.~\ref{rel_time_delay_Ip}.

\section{Conclusions}
\label{conc}

We have carried out an investigation of the relativistic regime of
tunnel-ionization with an emphasis on the role of the under-the-barrier
dynamics. In the quasistatic limit, the potential barrier of
relativistic tunneling can be defined in a gauge invariant manner by
means of an analysis of the physical energy operator. In contrast to the
nonrelativistic case, relativistic tunnel-ionization in the quasistatic
limit is modeled as tunneling through a potential barrier in an
additional magnetic field.  We showed that this problem can be reduced
to one-dimensional tunneling with a coordinate dependent energy.

The momentum distributions of the ionized electron wave packet at the
tunnel exit have been calculated using the SFA.  We showed that the
Lorentz force due to the magnetic field during the under-the-barrier
motion induces a momentum shift of the electron in the laser's
propagation direction.

In the second part of the paper, the problem of tunneling time delay has
been considered. Although, there is no well-defined time operator in
quantum mechanics, it is possible to infer information about the
tunneling time delay via tracing the peak of the wave packet, which
brings in the so-called Wigner time concept. The Wigner time formalism
was applied to nonrelativistic as well as to relativistic
tunnel-ionization. It was shown that the Wigner time formalism can be
simplified further for the tunnel-ionization process, due to the fact
that the quasiclassical trajectory starts at the entry point of the
barrier. In the nonrelativistic case, it was illustrated that the Wigner
time delay vanishes for the deep-tunneling regime when the potential
barrier at the tunneling exit can be approximated solely by its tangent
line. At larger laser field strengths, in the near-threshold-tunneling
regime of tunnel-ionization, the potential barrier is not linear in
coordinate at the tunnel exit. Consequently, the Wigner time delay is
preserved at far distances. Finally, our results were extended to the
relativistic regime. It was shown that the Wigner time delay is
characterized mainly by the nonrelativistic dynamics.


\bibliography{yakaboylu_bibliography}

\end{document}